\documentclass[hyper]{JHEP3}

\usepackage{epsfig}
\usepackage{latexsym}
\usepackage{amsfonts}
\usepackage{amsmath}
\usepackage{amsthm}
\usepackage{amssymb}
\usepackage{amsbsy}
\usepackage{multirow}
\usepackage{slashed}

\newcommand{\figref}[1]{Fig. \ref{#1}}
\newcommand{\secref}[1]{Sec. \ref{#1}}
\newcommand{\appref}[1]{Appendix \ref{#1}}
\newcommand{\tableref}[1]{Table \ref{#1}}

\def\({\left (}
\def\){\right )}

\def\cald         {{\cal D}}

\def\calm         {{\cal M}}
\def\caln         {{\cal N}}

\newsavebox{\uuunit}
\sbox{\uuunit}
    {\setlength{\unitlength}{0.825em}
     \begin{picture}(0.6,0.7)
        \thinlines
        \put(0,0){\line(1,0){0.5}}
        \put(0.15,0){\line(0,1){0.7}}
        \put(0.35,0){\line(0,1){0.8}}
       \multiput(0.3,0.8)(-0.04,-0.02){12}{\rule{0.5pt}{0.5pt}}
     \end {picture}}

\def\be{\begin{equation}}
\def\ee{\end{equation}}
\def\bea{\begin{eqnarray}}
\def\eea{\end{eqnarray}}

\newcommand{\beq}{\begin{eqnarray}}
\newcommand{\eeq}{\end{eqnarray}}

\def\a{\alpha}
\def\b{\beta}
\def\g{\gamma}
\def\G{\Gamma}
\def\d{\delta}
\def\e{\epsilon}

\def\l{\lambda}
\def\L{\Lambda}
\def\k{\kappa}
\def\f{\phi}

\def\m{\mu}
\def\n{\nu}
\def\o{\omega}

\def\p{\pi}
\def\r{\rho}

\def\s{\sigma}
\def\S{\Sigma}
\def\t{\theta}

\newcommand{\R}{\ensuremath{\mathbb{R}}}
\newcommand{\Z}{\ensuremath{\mathbb{Z}}}

\def\pa{\partial}
\def\T{\tau}

\def\to{\rightarrow}
\def\nonu{\nonumber \\{}}
\def\half{{1 \over 2}}
\def\sF{{{ F}\!\!\!\!\hskip.8pt\hbox{\raise1pt\hbox{/}}\,}}
\def\som{{{ \omega}\!\!\!\!\hskip.8pt\hbox{\raise1pt\hbox{/}}\,}}
\def\sJ{{{\rm J}\!\!\!\!\hskip.8pt\hbox{\raise1pt\hbox{/}}\,}}
\def\ve{\varepsilon}

\preprint{WITS-CTP-041\\KUL-TF-09/20}

\title{G\"odel space from wrapped M2-branes}
\author{T.S. Levi$^1$, J. Raeymaekers$^2$, D. Van den Bleeken$^3$, W. Van Herck$^4$, B. Vercnocke$^{4,5}$

\\

$^1$ Department of Physics and Astronomy,
University of British Columbia, \\
Vancouver, B.C. V6T 1Z1, Canada

\\

$^2$ Institute of Physics of the ASCR, v.v.i.\\
Na Slovance 2, 182 21 Prague 8, Czech Republic\\

$^3$ NHETC and Department of Physics and Astronomy, Rutgers University, \\
Piscataway, NJ 08855, USA \\

$^4$ Institute for Theoretical Physics,
K.U.Leuven\\
Celestijnenlaan 200D, B-3001 Leuven, Belgium\\

$^5$Jefferson Physical Laboratory, Harvard University,\\
Cambridge, MA 02138, USA

} \abstract{We show that M-theory admits a supersymmetric compactification to the G\"odel universe of the form G\"odel$_3\times$S$^2\times$CY$_3$. We interpret this geometry as coming from the backreaction of M2-branes wrapping the S$^2$ in an AdS$_3 \times$S$^2 \times$CY$_3$ flux compactification. In the black hole deconstruction proposal
similar states give rise to the entropy of a D4-D0 black hole. The system is effectively described by a three-dimensional theory consisting of an
axion-dilaton coupled to gravity with a negative cosmological
constant. Other embeddings of the three-dimensional theory imply similar
supersymmetric G\"odel compactifications of type IIA/IIB string
theory and F-theory.}

\begin{document}

\section{Introduction}
BPS states have played a major role in the successes of string
theory, from the understanding of black hole microstates to
nonperturbative checks of dualities. An interesting set of BPS states is that of supersymmetric
D-branes in an AdS$_q \times$S$^p$ background  (see e.g
\cite{Witten:1998xy,Bachas:2000ik,Simons:2004nm,Raeymaekers:2006np,Raeymaekers:2007vc}
and references therein). Such states are of interest for the
AdS/CFT correspondence in general. Furthermore, in the special
case where the background geometry corresponds to the near horizon of an
extremal black hole, string or ring, there are strong indications
that such BPS states, formed by wrapping branes around the S$^p$
part of the geometry correspond to black hole (string or ring)
microstates \cite{Gaiotto:2004pc,Gaiotto:2004ij,Denef:2007yt,Gimon:2007mha}.
The study of such
sphere-wrapping branes has so far been performed purely in the
probe approximation
\cite{Simons:2004nm,Gaiotto:2004pc,Gaiotto:2004ij,Das:2005za}.
There are however some indications that these branes strongly
backreact on the background geometry, and that some of their
properties can only be fully understood once these effects are
properly taken into account.

In this paper we take a first step at studying the fully
backreacted geometries corresponding to such wrapped branes. We
will specialise to the M-theory flux compactification
AdS$_3\times$S$^2\times$CY$_3$ and construct supergravity
solutions corresponding to M2-branes wrapped around the S$^2$.
Note however, that by taking the CY$_3$ to be T$^6$ or
K3$\times$T$^2$ and applying U-dualities these solutions can be
mapped to similar configurations in type IIA/IIB string theory or
F-theory.

We start our search for these solutions by noting that all the
dynamics can be captured by a reduction to three
dimensions and performing a consistent truncation to the fields of
interest. As we will discuss in detail, the problem can be brought
back to studying three-dimensional gravity with a negative
cosmological constant, coupled to an axion-dilaton system: \be
\frac{S_{3d}}{2\pi}=\frac{1}{l_{3}}\int
dx^3\,\sqrt{-g}\left(R+\frac{2}{l^2}-\frac{1}{2}\frac{\partial_\m
\tau\partial^\m \bar \tau}{\tau_2^2}\right)\,. \ee To the authors'
knowledge this three-dimensional theory has never before been studied
in the literature. This is somewhat surprising as three-dimensional 
gravity with a negative cosmological constant is a
surprisingly rich gravitational theory that is well explored and
remains the subject of present investigations (see
\cite{Witten:2007kt,Li:2008dq} and references therein).
Furthermore, the above theory without a cosmological constant was
the subject of the classic paper \cite{Greene:1989ya}, and is very
closely related to F-theory.

Due to its embedding in eleven-dimensional supergravity the above bosonic action
is naturally completed into a supersymmetric theory. We will show
that this theory has 1/2-BPS solutions that are all locally
G\"odel space\footnote{G\"odel's original spacetime was
four-dimensional, but it is nothing but the direct product of a
non-trivial three-dimensional spacetime with a space-like line. It is
this three-dimensional spacetime that we will refer to as G\"odel
space.}:
\bea
ds^2_3&=&\frac{l^2}{4}\left(-(dt+\frac{3}{2}\frac{dx}{y})^2+\frac{3}{2}\frac{dx^2+dy^2}{y^2}\right) ,\\
\tau&=&x+iy . \eea The full eleven-dimensional solution can be read
off by substituting this metric in formula (\ref{11Dreduction})
below. G\"odel space \cite{Godel:1949ga} has a long history, and
this work provides a new supersymmetric embedding into
string/M-theory. For an earlier example see \cite{Israel:2003cx}.
More precisely our work shows that M-theory has a compactification
of the form G\"odel$_3\times$S$^2\times$CY$_3$ that preserves 4
out of the 32 supersymmetries. For other embeddings of spaces with
closed timelike curves in string theory, see e.g.
\cite{Gauntlett:2002nw,Compere:2008cw}.

As the real part of the scalar field is an axion which is dual to the
gauge field that is sourced by the sphere-wrapping M2-branes, the
G\"odel universe carries this charge. The field configuration of
global G\"odel space corresponds to an infinite amount of
M2-charge localised in a point on the boundary, $y=\infty$.

G\"odel space suffers from closed timelike curves (CTCs). We
study some simple domainwall configurations, made out of smeared
M2-branes, that allow us to glue G\"odel space to AdS$_3$. It was our hope that this would eliminate the CTCs as it does for a similar system dubbed the `hypertube' \cite{Gimon:2004if}. Unfortunately, at present we have not been able to find a global solution that fully eliminates CTCs. We remain optimistic that a future treatment, either with another patching or a smooth resolution of the patching such as  was found for the hypertube  \cite{Berglund:2005vb,Bena:2005va} will resolve this issue. Nevertheless, the glued geometries we have found seem very interesting from the point of view of holography.

The paper is organized as follows. In the next section we review
and discuss the supersymmetric properties of the wrapped M2-brane
states that motivate our study. In \secref{secto3d} we then detail
our flux compactification of M-theory to three dimensions and how
solutions sourced by the wrapped branes correspond to effectively
three-dimensional geometries with a non-trivial axion-dilaton profile.
In \secref{section-godel}, we solve the equations of motion of the
three-dimensional theory and show these solutions are supersymmetric.
\secref{sec-glue} covers how the G\"odel space can be
supersymmetrically glued to AdS space through the introduction of
an appropriate domainwall. \secref{sec-con} presents some
discussion and suggestions for future directions. For the reader's
convenience we have provided some appendices containing extra
technical details. \appref{app-duality} describes in some more
detail the U-dualities that link our solutions to similar ones in
dual frames. The supersymmetry of the G\"odel solutions in eleven- and
five-dimensional supergravity is carefully shown in full detail in
\appref{app-susy} and \appref{app-5Dsusy}. In \appref{IsJuncAp} we
quickly review the Israel matching conditions using a rather
simple approach that clarifies the generalisation to more general
field theories.

\section{Sphere-wrapping M2-branes in the probe approximation}\label{sec-probe}
Before embarking on the construction of backreacted solutions of
S$^2$-wrapping M2-branes, let us review what is known about these
BPS objects in the probe approximation
\cite{Simons:2004nm,Gaiotto:2004pc,Das:2005za}. These properties
will be useful when comparing  with the backreacted solutions in
\secref{section-godel}.

Our starting background is M-theory compactified on AdS$_3 \times $S$^2 \times
$CY$_3$, and we will assume that the anti-de-Sitter factor is global
AdS$_3$ and not a local solution such as a BTZ black hole. Such a
background arises e.g. in a certain limit of a D6-anti-D6
configuration when lifted to M-theory \cite{Denef:2007yt,deBoer:2008fk}. The
AdS$_3$ part of the metric is \be
ds_3^2=l^2\left[-\cosh^2\r\,d\sigma^2+d\r^2+
\sinh^2\r\,d\psi^2\right]\, \label{globalmetr} \ee and the bosonic symmetry group of the eleven-dimensional background is
$SL(2,\R )_L \times SU(2)_L \times SL(2,\R )_R$.

We add to this background a probe M2-brane (or anti-M2-brane) wrapped on S$^2$, which behaves as
a massive point particle in AdS$_3$. As can be seen from (\ref{globalmetr}), we cannot place
a static particle  (with respect to the global time $\s$) at finite $\r$, as it experiences a gravitational potential and will fall towards $\r = 0$.
A spinning particle however, obeying $\psi = \s + {\rm constant} $, can stay at any fixed constant radius $\r = \r_0$. The $\r_0$-dependent momentum conjugate to
$\psi$ determines the D0-brane charge after compactification on a circle  to type IIA in ten dimensions.
Such spinning M2-branes are BPS states
and are the objects we will study, see \figref{probebr}.
\FIGURE{
\includegraphics[scale=0.6]{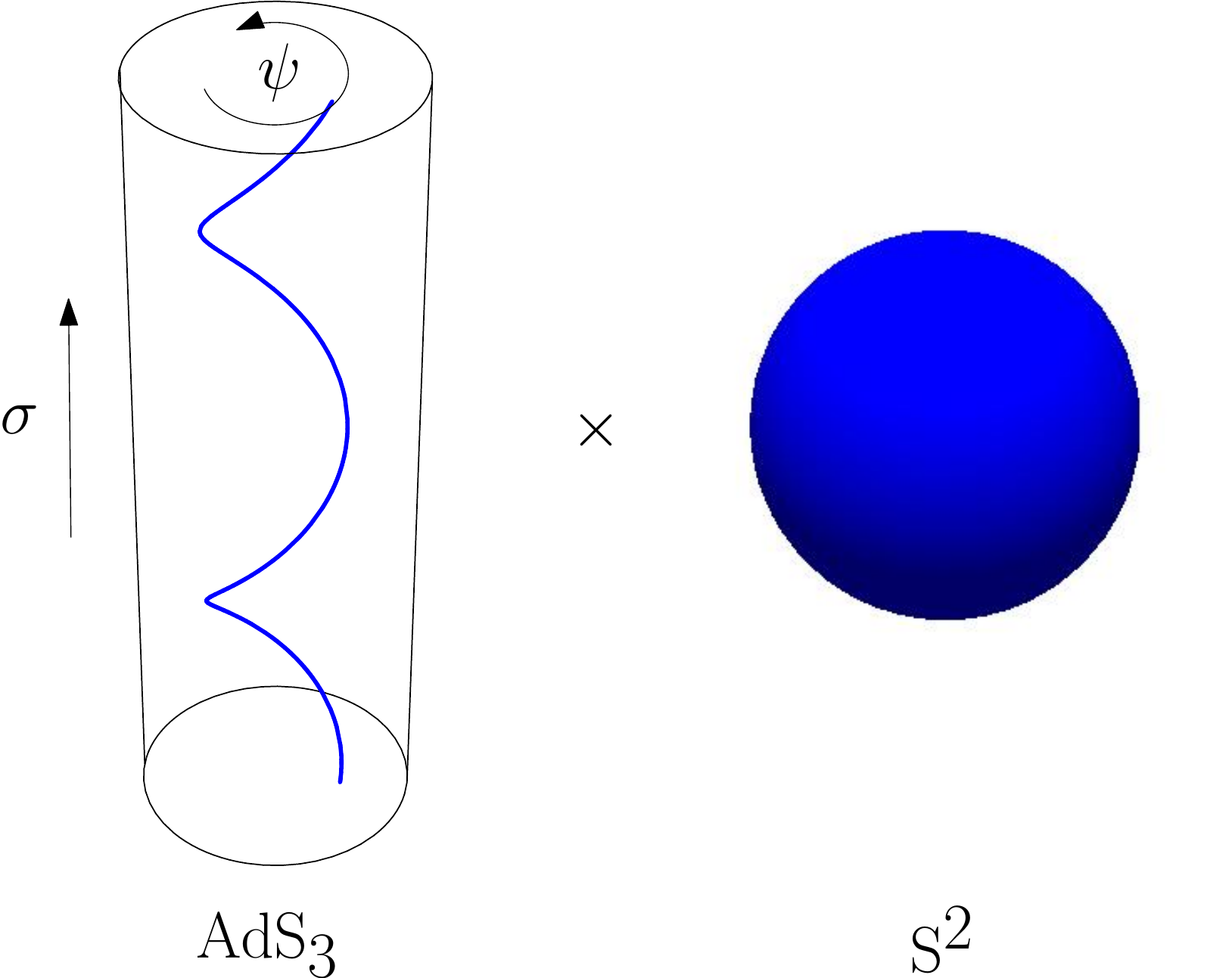}\caption{The BPS M2-brane wraps the S$^2$ and has a helical worldline, $\psi=\sigma+$constant,  in the external AdS$_3$ space.}\label{probebr}}

It will be convenient to introduce a new time coordinate with
respect to which the M2-brane is at rest; if we define $ \s =t/2,\
\psi = t/2 - \varphi$, our probe brane is static with respect to
$t$. The metric in these coordinates becomes \be ds^2 = {l^2 \over
4} \left[ - ( d t + 2 \sinh \r\, d \varphi)^2 + 4 d \r^2 + \sinh^2
2 \r\,  d \varphi^2\right] \label{timet} . \ee Note that the AdS$_3$
metric is now written as a timelike fibration over Euclidean
AdS$_2$. The new time coordinate also has a useful
group-theoretical interpretation: it  is the time coordinate
adapted to the $l_0$ generator of $SL(2,\R )_L$: $l_0 = \pa /
\pa_t$. In the presence of the probe brane, the symmetry group is
reduced to the generators that commute with $l_0$, leaving $U(1)_L
\times SU(2)_L \times SL(2,\R )_R$. The  $l_0$ Noether charge is the
Hamiltonian with respect to $t$, and one finds the dispersion
relation \be L_0 = { l \over 2} T_{M_2} {\rm Vol}_{S^2} \equiv Z
\label{Noethermain}, \ee where $L_0$ is the eigenvalue of $l_0$,
$T_{M_2}=2 \pi \ell_M ^{-3}$ is the tension of the M2-brane and
$l_M$ is the eleven-dimensional Planck length. We have used the symbol
$Z$ to denote that the right-hand side is a topological charge
proportional to the winding number (which we have taken to be
unity) of the brane around S$^2$; we will see below that $Z$
enters as a central extension in the worldvolume superalgebra.

The eleven-dimensional background preserves 8 fermionic symmetries which combine with
$SL(2,\R )_L \times SU(2)_L $
to form the supergroup
$SU(1,1|2)_L$.
The corresponding Killing spinors can be labelled as $g^{\; \a
a}_{m\over 2}$ where the indices $m, \a, a$ each take the values
$\pm 1$. The indices $m$ and $\a$ are  $SL(2,\R )_L$ and $SU(2)_L$
doublet indices respectively. The explicit form of the Killing
spinors can be found in
(\ref{Killspinorbasis},\ \ref{constspinors}).
 From the Killing
vectors and Killing spinors one can compute the isometry
supergroup of the supergravity background using the method
developed in \cite{Gauntlett:1998kc}. This gives the
fermionic part of the  of $SU(1,1|2)_L$ superalgebra: \bea \{g_{m \over 2}^{\a a},
g_{m \over 2}^{\b b}\} &=& \e^{\a\b}\e^{ab} l_m , \nonu \{g_{m \over
2}^{\a a}, g_{-{m \over 2}}^{\b b} \} &=& \e^{\a\b}\e^{ab} l_0 + m
\e^{ab} T^{\a\b} , \eea where $T^{\a\b}$ is an $SU(2)_L$ tensor
given in (\ref{Ttensor}).

The (anti-)M2-brane probes preserve 4 out of 8 Killing spinors, as one can verify
using standard methods \cite{Bergshoeff:1987cm,Bergshoeff:1997kr}.
In terms of our basis $g^{\; \a a}_{m\over 2}$, the preserved
Killing spinors are \bea g_{1 \over 2}^{\a +},\ g_{-{1 \over
2}}^{\a -} & \qquad &\text{M2-brane} ,\nonu g_{1 \over 2}^{\a -},\
g_{-{1 \over 2}}^{\a +} & \qquad &\text{anti-M2-brane} .
\label{probeKSmain} \eea The dispersion relation (\ref{Noethermain}) can be
viewed as a BPS condition, with $Z$ playing the role of a central charge.
The extended superalgebra can be computed
using the methods of \cite{HackettJones:2003vz}. The result is
\bea \{g_{m \over 2}^{\a a}, g_{m \over 2}^{\b b}\} &=&
\e^{\a\b}\e^{ab} l_m ,  \nonu \{g_{m \over 2}^{\a a}, g_{-{m \over
2}}^{\b b} \} &=& \e^{\a\b}\e^{ab} l_0 + m \e^{ab} T^{\a\b} \mp m
Z \e^{\a\b} \s_1^{ab} \label{centralextmain} , \eea where the (plus) minus sign corresponds
to a (anti-)  brane. This is a centrally extended version of $SU(1,1|2)_L$.  In the quantum theory, chiral primaries of
the extended algebra saturate a BPS bound
\be
L_0 \geq j +  Z .
\ee
From (\ref{Noethermain}) we  see that our probes saturate this BPS
bound for $j=0$ and that the preserved supersymmetries (\ref{probeKSmain})
are  a consequence of  the extended superalgebra.

The fact that the wrapped M2-brane is a chiral primary state in a
centrally extended version of $SU(1,1|2)_L$ leads to a useful
observation. The asymptotic symmetry algebra of the AdS$_3 \times
$S$^2$ background, which coincides with  the superconformal algebra
of the dual CFT (see \cite{Gaiotto:2006wm,deBoer:2006vg} for a
detailed description of this algebra), does not contain the
central charge term in (\ref{centralextmain}). Therefore we can
expect that the backreacted geometry of the wrapped M2-probes is
no longer asymptotically AdS$_3 \times $S$^2$.  Another reason for
expecting a large backreaction destroying the asymptotics of the
background is the fact that we are dealing with charged
codimension-two objects, which produce long-range fields. Indeed,
we will find that, after backreaction, the AdS$_3$ factor is to be
replaced with G\"{o}del space, which is not an  asymptotically
AdS$_3$ geometry. We will construct such  backreacted solutions
and show that they  preserve the same bosonic  $U(1)_L \times
SU(2)_L \times SL(2,\R )_R$  symmetries and the same supersymmetries
(\ref{probeKSmain}) as the probe.

\section{Effective three-dimensional description}\label{secto3d}
The arguments of the last section lead us to search for supersymmetric solutions where the fields that are sourced by the sphere-wrapped M2-branes under consideration are turned on. We will show that there is a consistent reduction on the S$^2 \times $CY$_3$ that includes these fields and results in the three-dimensional action%
\be \label{3daction}
\frac{S_{3d}}{2\pi}=\frac{1}{l_{3}}\int dx^3\,\sqrt{-g}\left(R+\frac{2}{l^2}-(\m -1)\frac{\partial_\m \tau\partial^\m \bar \tau}{\tau_2^2}\right)\, ,
\ee%
which will be the starting point for the analysis in the rest of the paper.

This system describes three-dimensional gravity with a negative cosmological constant $\Lambda = -1/l^2$, coupled to a complex scalar $\tau = \tau_1 + i\tau_2$, which has a typical axion-dilaton type kinetic term. In the $l\rightarrow\infty$ limit this action is that of \cite{Greene:1989ya}. The eleven-dimensional origin of the scalar $\tau_1$ is as the Hodge dual of the electric field sourced by the M2-branes, while $\tau_2$ is the dynamic Calabi-Yau volume.
We have introduced a coupling constant $\mu$ in front of the axion-dilaton action. Positive energy requires $\m >1$, and we will see that the M-theory reduction fixes $\m = 3/2$, which is the value we will be interested in.
Nevertheless, we will be  able to construct solutions for more general $\mu$, although it is not clear whether these allow for a supersymmetric embedding.

We show in more detail how the above three-dimensional system can be derived from M-theory. We begin with the bosonic part of the eleven-dimensional supergravity (M-theory) action. In our conventions it is given by
\be
\frac{S_{M}}{2\pi}=\frac{1}{l_{M}^9}\int d^{11}x\sqrt{-g}R-\frac{1}{2l_{M}^3}\int F_4\wedge\star F_4+\frac{1}{6}\int A_3\wedge F_4\wedge F_4, \label{Maction}
\ee
where $l_M$ is the eleven-dimensional Planck length and $F_4=dA_3$. We will also use the M2-brane action
\be
\frac{S_{M2}}{2\pi}=-\frac{1}{l_M^3}\int_{\S}d^3\xi\sqrt{-g}+\int_{\S}A_3\,.
\ee
We seek solutions that are sourced by M2-branes that wrap the two-sphere in the  flux compactification of M-theory on S$^2\times$CY$_3$. We make a consistent reduction for such solutions using the ansatz 
\bea
ds^2_{11}&=&\tau_2^{-2/3}\left(ds^2_3+\frac{l^2}{4}d s^2_{S^2}\right)+l_M^2\tau_2^{1/3}d s^2_{CY_3}\,, \label{11Dreduction}\\
F_4&=&-\frac{\star_3 d\tau_1}{l_3 \tau_2^2}\wedge\o_{2}+\frac{2\pi\,l}{l_M}\o_{2}\wedge J_{CY_3}\,.\label{F4}
\eea%
The Hodge dual $\star_3$ is taken with respect to $ds^2_3$. We have furthermore parameterized the dynamic Calabi-Yau volume with a scalar $\tau_2$, so that in the notation above both the S$^2$ and the CY$_3$ part of
the metric have fixed unit volume. The volume form on the unit S$^2$ is $\o_2$, and $J_{CY_3}$ is the K\"ahler form on the unit volume Calabi-Yau.

In this ansatz we allow two contributions to the M-theory gauge field $F_4$: the part along $\omega_2\wedge J_{CY_3}$ describes the flux needed to support the background, while the first term in \eqref{F4} is the one sourced by
sphere-wrapped M2-branes. We describe this component in terms of a real scalar field $\tau_1$, as this is more natural from the three-dimensional point of view.   It is related to a $U(1)$-potential $A$ by
three-dimensional Hodge duality:
\beq
dA = -\frac{\star_3 d\tau_1}{l_3\,\tau_2^2}\,.\label{Fvstau1}
\eeq
The three-dimensional Planck length $l_3$ and the AdS-radius $l$ are related to the eleven-dimensional Planck length $l_M$ as
\beq%
l_M^3=\p\,l_3\,l^2\,.
\eeq%
Finally, it is natural in three dimensions to combine the dualized electric field $\tau_1$ and the Calabi-Yau volume $\tau_2$ into one complex scalar $\tau$ as:
\beq
\tau = \tau_1 + i\tau_2\,.
\eeq
One can then check that for an ansatz of the form specified above, the eleven-dimensional equations of motion become equivalent to those of the three-dimensional action \eqref{3daction}. In such a reduction the sphere-wrapping M2-branes can be described as charged particles:
\beq
\frac{S_{M2}}{2\pi}&=&-\frac{1}{l_3}\int d\xi \frac{\sqrt{-g}}{\tau_2}+\int A\,.\label{sourceterms}
\eeq%

\noindent We would like to point out a few interesting facts about this reduction:%
\begin{itemize}
\item We were able to reduce assuming a constant radius for the S$^2$ because none of the other five-dimensional fields or the M2-brane source terms couple to the volume of the sphere.
The latter is a general property of codimension-two branes: if one considers a codimension-two brane wrapped on a compact manifold, and reduces over this manifold, the coupling of the BI action to the volume
modulus is balanced by the Weyl transformation needed to go to the lower-dimensional Einstein frame.

\item Although we will be mainly interested in the M-theory origin
of the three-dimensional system  (\ref{3daction}) discussed above, solutions to
(\ref{3daction}) can of course  be embedded in any higher
dimensional theory that allows (\ref{3daction}) as a consistent
truncation. Of special physical interest are embeddings in a type
IIB string theory on S$^3 \times \calm_4$ (with $\calm_4$ either
$K_3$ or $T_4$), which is the near horizon limit of the D1-D5
system. In the probe approximation, branes wrapping the S$^3$ have
been conjectured to account for the entropy of the Strominger-Vafa
black hole in a similar manner as the S$^2$-wrapping M2-branes in
the  M-theory frame did for the current setup
\cite{Raeymaekers:2007ga}. Performing an S-duality, one obtains
another  interesting duality frame where only NS sector fields are
excited and which could be the starting point for a sigma-model
description. In another duality frame of interest, our system
describes the backreaction of D7-branes wrapped on  S$^3 \times
\calm_4$. In this frame, the field $\T$ is the standard
axion-dilaton of type IIB. Our configurations can then be viewed
as nontrivial solutions  of F-Theory. In \appref{app-duality} we
discuss how our solutions can be embedded into different higher-dimensional 
theories by showing the explicit U-duality chain. An
overview is presented in \tableref{interestingframes}.
\begin{table}
\begin{center}
\begin{tabular}{|c|c|c|}\hline
theory & background branes & source branes\\ \hline
$M$ on  S$^2 \times T^6$ & $M5$'s & $M2$ on S$^2$ \\
$IIB$ on  S$^3 \times T^4$ & $D3-D3$ & $D3$ on S$^3$ \\
$IIB$ on  S$^3 \times T^4$ & $D1-D5$ & $D5$ on S$^3 \times T^2$ \\
$IIB$ on  S$^3 \times T^4$ & $F1-NS5$ & $NS5$ on S$^3 \times T^2$ \\
$F$ on  S$^3 \times T^6$ & $D3-D3$ & $D7$ on S$^3\times T^4$ \\ \hline
\end{tabular}
\end{center}\caption{Embeddings of the three-dimensional axion-dilaton system in various higher dimensional theories related by U-duality.
The axion-dilaton $\T$ is in each case a modulus of the toroidal compactification.
The background branes produce a near horizon AdS$_3$ flux compactification, and the source branes couple to $\T$.}\label{interestingframes}
\end{table}

\item{Instead of compactifying to three dimensions, one can instead consider the five-dimensional theory obtained by reduction on the Calabi-Yau alone. This reduction gives us the action of $\mathcal{N}=1$ supergravity in five dimensions. The complex scalar $\tau$ is then part of the universal hypermultiplet. To account for the charge and backreaction of probe branes wrapped on S$^2$  we must seek solutions with non-trivial hyperscalars turned on. There is an extensive literature on the general framework of finding solutions in this situation \cite{Bellorin:2006yr,Huebscher:2006mr,Strominger:1997eb,Behrndt:2000zh}, though few specific examples are known.
Throughout the rest of the paper we will focus on the M-theory language and the three-dimensional action which are more straightforward. However, it is important to note that our general ansatz and
solutions are non-trivial solutions of the five-dimensional theory. To the extent of the authors' knowledge these solutions are the first non-singular examples of such type. It might be interesting to make the
connection to the geometric language of \cite{Bellorin:2006yr} more precise.
}

\end{itemize}

\section{Holomorphic BPS solutions} \label{section-godel}
In this section we study stationary solutions to the action \eqref{3daction} and display an ansatz that will lead to BPS solutions. Before we proceed with solving the equations of motion, we remark that we are looking at a special system. In three dimensions,  the M2-branes wrapped on the sphere correspond to particles and have a one-dimensional worldvolume.
However, charged codimension-two objects exhibit non-generic behaviour. Unlike many other (p-) brane solutions, the fields they source depend not only on one radial variable, but on the two directions of the transverse space.
This was shown for the flat space analog of our system for instance in the case of cosmic strings in four dimensions \cite{Greene:1989ya} and later applied to seven-branes in type IIB supergravity \cite{Gibbons:1995vg}. Our work generalises such systems by the addition of  a negative cosmological constant.

\paragraph{Ansatz:}
The equations of motion one obtains from \eqref{3daction} are%
\bea
R_{\a\b} &+&\frac{2}{l^2}g_{\a\b}=(\m - 1)\frac{\partial_{(\a}\tau\partial_{\beta)}\bar\tau}{\tau_2^2} \, ,\label{einsteineqns}\\
\partial_\alpha\left(\sqrt{-g}g^{\a\b}\partial_\b\tau\right) &+& i\sqrt{-g}g^{\a\b}\frac{\partial_\a\tau\partial_\b\tau}{\tau_2}=0 .\label{scalareqns}
\eea
It is important to note that, due to the specific `non-standard' kinetic term for the scalar $\tau$, equation \eqref{scalareqns} --- from varying with respect to $\bar \T$ --- only features derivatives
of $\tau$ and not of $\bar \tau$. Furthermore, since the probe branes we started from were time-independent for some specific timelike coordinate $t$ (see (\ref{timet})),
we seek stationary solutions: $\partial_t\tau=0$. As was shown in \cite{Greene:1989ya} for the flat space analog of our system ($l=\infty$), these two facts combined imply
that the scalar equation of motion can be solved by choosing $\tau$ to be (anti)-holomorphic in the complex coordinate naturally made up of the remaining two spatial coordinates. One can see from \eqref{scalareqns}
that this remains true even if the full three-dimensional metric is not flat but when the spatial part of $\sqrt{g}g^{\a\b}$ consists of constants.

Any stationary metric in three dimensions can be written as
\be
ds^2_{3d}=\frac{l^2}{4}\left(-e^{2\lambda}(dt+\chi)^2+e^{2\phi}ds^2_{F}\right) \, .\label{metricansatz}
\ee
Here $ds^2_F$ is a flat metric on the two spatial directions, $\phi$ and $\lambda$ are functions of the spatial coordinates only and $\chi$ is a one-form on the spatial part.

The assumption that for the spatial directions $\sqrt{g} g^{\a \b}=$constant is equivalent to requiring $\lambda=0$.
Choosing complex coordinates on $ds^2_F$, our metric ansatz takes the form
\bea
ds^2_{3d}&=&\frac{l^2}{4}\left(-(dt+\chi)^2+e^{2\phi(z,\bar z)}dzd\bar z\right)\label{stationaryLiouville} \, .
\eea
There is still some gauge freedom left in this ansatz. We can make a conformal transformation $z\rightarrow g(z)$ which does not alter the form of the metric but
sends
\be \phi(z,\bar z)\rightarrow\phi(z,\bar z)+\frac{1}{2}\log \pa g(z)+\frac{1}{2}\log\bar \pa \bar g(\bar z)\label{conftransf}.\ee
There is also a local shift symmetry
\be t\rightarrow t-f(z,\bar z), \qquad \chi\rightarrow\chi+d f.\label{shiftsymm}\ee

\paragraph{Equations of motion:}

With the metric ansatz (\ref{stationaryLiouville}), the $\T$ equation of motion (\ref{scalareqns})  reduces to
\be
\pa \bar \pa \T + i {\pa  \T \bar \pa \T  \over \T_2} = 0 \label{taueqn}.
\ee
It can be solved by taking $\T$ to be an arbitrary holomorphic or antiholomorphic function, and we will see in the next section that this leads to supersymmetric solutions.

Using our ansatz (\ref{stationaryLiouville}), the Einstein equations \eqref{einsteineqns} can be written as 
\bea
d\chi&=&\frac{ie^{2\phi}\,dz\wedge d\bar z}{2}\label{chieqn} \, ,\\
\partial\bar\partial\phi-\frac{e^{2\phi}}{4}&=&- (\m - 1)\frac{\partial\tau\bar\partial\bar\tau}{4 \T_2^2} .\label{liouvilleeqn}
\eea
The equation for  $\phi$ is the Liouville equation with a source term provided by $\T$.

Before we solve these equations, it is instructive to note the differences with the flat space limit of \cite{Greene:1989ya}. In that flat space scenario, the equation for the conformal factor $\phi$ is a Poisson equation with source where here we found a sourced Liouville equation. Another important difference is the topology of the spatial base manifold. In the presence of a cosmological constant the spatial base manifold  has the conformal structure and topology of the disk,  as opposed to the Minkowski case where the topology and conformal structure underlying the equations is that of the Riemann sphere. In our case the equations (\ref{chieqn}),(\ref{liouvilleeqn}) still have an elegant solution, but it is not straightforward to construct `stringy' solutions where $\T$ has nontrivial $SL(2,\Z )$ monodromies as was done in \cite{Greene:1989ya}.

\subsection*{Constant axion-dilaton: AdS$_3$ }
When $\T$ is constant, we are in the pure gravity case and the metric (\ref{stationaryLiouville}) describes local AdS$_3$, written as a timelike fibration over
Euclidean AdS$_2$.  We illustrate how AdS$_3$ can be written in this form, and  introduce
two coordinate systems that will appear later on.

The general solution to the Liouville equation (\ref{liouvilleeqn}) without source and to the equation for the one-form $\chi$  (\ref{chieqn}) is\footnote{For notational simplicity we are a bit sloppy in distinguishing between the holomorphic partial derivative and the corresponding Dolbeault operator, denoting both with $\partial$. We trust the reader to distinguish between them by checking if the result is a scalar or differential form.}
\bea
e^{2 \phi} &=& {4 \partial g \bar \partial \bar g \over (1- g \bar g)^2}\label{Liouvillesolvac} \, , \\
\chi  &=&  2 \mathrm{Im} \pa \phi  + d f \label{chisolvac} \, ,
\eea
where $g (z)$ is an  arbitrary holomorphic function and $f (z, \bar z)$ is an arbitrary real function. These arbitrary functions reflect the conformal invariance (\ref{conftransf}) and the shift symmetry (\ref{shiftsymm}) of our
ansatz.
The resulting metric is locally AdS$_3$. To see this more explicitly,
one can make the following coordinate transformation to global AdS$_3$ coordinates $(\s, \rho, \psi)$:
\bea
\s &=& {t + f\over 2} , \nonu
\rho &=& {\rm arctanh }\  | g| , \nonu
\psi &=& -\arg ( g ) + {t + f \over 2} \label{transfoglobal} ,
\eea
in terms of which one obtains the standard global AdS$_3$ metric (\ref{globalmetr}). As discussed in  \secref{sec-probe}, the AdS$_3$ vacuum preserves
8 Killing spinors in the theory under consideration. These were
labelled as
 $g^{\; \a a}_{m\over 2}$ and correspond to the fermionic generators of $SU(1,1|2)_L$. They are computed in \appref{app-susy} and are given by
\be g_{m \over 2}^{\a a} = {\sqrt{l} \over \sqrt{2} \T_2^{1/6}}  \;
e^{ {i m\over 2}( t + f)} e^{ - {i \a \varphi\over 2} }
e^{ {i \t \over 2}  \g^{\hat \varphi}} g_{0\; {m \over 2}}^{\; \a
a} \label{KSmain} , \ee where the $g_{0\; {m \over 2}}^{\; \a a}$ are suitably
chosen constant spinors given in (\ref{constspinors}).

Two coordinate systems will be of use later.
In the first coordinate system, which we call `disk coordinates' $({t}, z, \bar z)$, the spatial base is the Poincar\'e  disk.
We take $z, \bar z$ to  range over the unit disk in the complex plane, $|z|<1$, and choose $g (z) = z,\ f = 0$. Putting $z= r e^{i  \varphi} $   we find
\beq%
ds^2=\frac{l^2}{4}\left[-\left(d{t}+\frac{2{r}^2}{1-{r}^2}d{\varphi}\right)^2+\frac{4}{(1-{r}^2)^2}(d{r}^2+{r}^2 d {\varphi}^2)\right].\label{diskmain}
\eeq%
This coordinate system covers AdS$_3$ globally as one can check using (\ref{transfoglobal}). It is related to the coordinate system (\ref{timet}) in \secref{sec-probe}
by a redefinition of the radial coordinate $r = \tanh \r$.  The boundary of AdS$_3$ is a cylinder formed by the unit circle  and the time coordinate.

In another coordinate system, which we will call `upper half
plane coordinates' $(t ,w,\bar w)$, the base is the hyperbolic
plane. We take coordinates $w, \bar w$ ranging over the upper half
plane, $\mathrm{Im} \, w>0$, and take $ g(w) = (w - i)/(w + i),\ f =0 $. Putting $w =
x + i y$, the metric becomes \be ds^2 = {l^2 \over 4} \left[ -
\left( d t + {dx \over y}\right)^2 + { dx^2 + dy^2 \over
y^2}\right]. \label{UHPmain}\ee This is again a global coordinate system\footnote{We should warn the reader that the $t$ coordinate in (\ref{UHPmain}) differs from the one in (\ref{diskmain}) by a shift transformation: $t\rightarrow t+2\arg(1-z)$.}, and
the spatial  boundary now consists of the real line and the point
$w = i \infty$.

\subsection*{Holomorphic axion-dilaton solutions: G\"{o}del space}

We explore non-trivial solutions to the
equations (\ref{chieqn},\ref{liouvilleeqn}) where the axion-dilaton field is  nonconstant. We will construct supersymmetric solutions which
should be seen as the backreacted geometries due to wrapped M2-brane sources.  The class of solutions we find, has brane sources on the boundary
and its local geometry is that of the three-dimensional G\"{o}del universe. In fact, the simplest solution is global G\"{o}del space,  thus providing a new supersymmetric embedding of the G\"{o}del universe in string/M-theory.

\paragraph{Solving the equations:}

We can solve the $\T$ equation (\ref{taueqn}) by
taking  $\T$ to be a holomorphic function:
\be \T = \T(w) \label{tausol}. \ee
As we will show below, this will lead to 1/2-BPS solutions\footnote{We could also take $\T$ to be antiholomorphic,
which would correspond to replacing brane sources with antibranes and would preserve different supersymmetries.}.

Next we turn to the equation for the one-form $\chi$  \eqref{chieqn}.  We see that it is solved by a simple modification of (\ref{chisolvac}):
\be
\chi = 2 \mathrm{Im} \left( \pa \phi + ( 1 - \m) \pa \ln \T_2 \right) + d f \label{chisol}.
\ee
Again, $f$ is an arbitrary real function reflecting the shift symmetry (\ref{shiftsymm}).

It remains  to solve the Liouville equation (\ref{liouvilleeqn}) for the conformal factor $e^{2 \phi}$ in the presence of the source term. We write the equation as
\be
\cald e^{2 \f} \equiv \left( \pa \bar \pa \ln - \half  \right) e^{2 \f} = - (\m - 1) {\pa \T \bar \pa \bar \T \over 2 \T_2^2 } \, .%
\ee The source term on the right-hand side is quite special in that it is an eigenfunction
of the the non-linear differential operator $\cald =\pa \bar \pa
\ln - \half $. Indeed, we can write it as \be
- (\m -1 ) {\pa \T \bar \pa \bar \T \over 2 \T_2^2} = \cald \left( \m {\pa \T \bar \pa \bar \T \over  \T_2^2 }\right) .
\ee
Therefore a solution to the equation is given by
\be e^{2 \phi } = \m {\pa \T \bar \pa \bar \T \over  \T_2^2 }\label{phisol} . \ee
This solution is a special case of those obtained in \cite{Semenov:2008}. Let us discuss the uniqueness of our solution. As far as the authors know, there does not exist a proof of uniqueness of the solution (\ref{phisol}) in the literature.
Nevertheless, since we are working in three-dimensional gravity, we know that a given energy-momentum tensor completely determines the local geometry, so that other
solutions to (\ref{liouvilleeqn}) (if any) must lead to a locally equivalent metric. Boundary conditions can then provide the global structure. However as we will see there seems to be a unique simply connected
and geodesically complete solution. In this solution  the spatial base, parameterized by $(w, \bar w)$, has  the conformal structure and topology of the disk, as in the case of global AdS$_3$.
In this paper we will not explicitly consider the interesting generalization of taking the base to be a quotient of this disk  and $\T$ to have nontrivial $SL(2,\Z )$ monodromies. We hope to return to this in the future.

\paragraph{Supersymmetry properties:}

Let us discuss the supersymmetry properties of the solutions (\ref{tausol},\ref{chisol},\ref{phisol}) when embedded in five- or eleven-dimensional supergravity.
The bosonic symmetry group  is broken
to $U(1)_L \times SU(2)_L \times SL(2,\R )_R$. In Appendix \ref{app-susy}, we show that
the backgrounds with nonconstant $\T$ are supersymmetric and preserve half of the supersymmetries preserved in the constant $\T$ case.
The preserved Killing spinors are precisely of the form (\ref{KSmain}) (with $\T_2$ no longer constant),
subjected to an additional projection condition leaving one half of the supersymmetries preserved by AdS$_3$. In terms of our  basis (\ref{KSmain}) of the $SU(1,1|2)$
superalgebra, the
Killing spinors preserved after turning on a holomorphic $\T$ are\footnote{If we had turned on antiholomorphic $\T$, we would have preserved the other half of the supersymmetries,
i.e. $g_{1 \over 2}^{\a -},\ g_{-{1 \over 2}}^{\a +}$.}
 $$g_{1 \over 2}^{\a +},\ g_{-{1 \over 2}}^{\a -}.$$
This is in precise agreement with the  analysis (\ref{probeKSmain}) in the probe approximation, and confirms our interpretation that our solutions represent
backreacted wrapped M2-branes. As a
further check on this interpretation, one can verify that placing a wrapped M2-brane probe in our nontrivial background does not break any further supersymmetries,
as is expected from standard D-brane lore, by studying the $\kappa$-symmetry conditions.

\paragraph{The G\"odel solution:}
As discussed above, we can take $\T$ to be any single-valued (possibly multiple-to-one) meromorphic function from the upper half plane to itself. The simplest case, which we will study in the remainder of
this work, is to take
\be
\T( w) = w\label{eq:tauw}
\ee
which is one-to-one and has a first order pole on the boundary at $w = i \infty$. More general multiple-to-one maps can be locally brought into this form by a conformal transformation.
Choosing $f=0$ in (\ref{chisol}) and defining $w = x + i y$ we obtain the metric%
\beq%
ds^2 = { l^2 \over 4} \left[ - (dt+ \mu {dx \over y})^2 + \mu {d x^2 + d y^2 \over y^2} \right] \, \label{godelUHPmain} .
\eeq%
Rescaling $t \to \mu t$ one sees that this is the metric of timelike warped AdS (see e.g. \cite{Bengtsson:2005zj}).

For $\m >1$, including the case of interest $\m = 3/2$,
the timelike fiber is stretched, and
the space
is known to be the G\"{o}del geometry \cite{Rooman:1998xf}. G\"{o}del's original solution \cite{Godel:1949ga} corresponds to setting $\mu = 2$. This is one of the first examples of a three-dimensional
supersymmetric G\"odel space in the literature, see also
\cite{Israel:2003cx}.

Formally one could also take $\m <1$, in which case the timelike fibre is squashed  with respect to pure AdS$_3$.
This space has no closed timelike curves \cite{Bengtsson:2005zj} and also appears as a solution to topologically massive gravity \cite{Anninos:2008fx}.
However it  arises from an unphysical matter source:  as we can see from  (\ref{3daction}), it requires a `ghost' axion-dilaton with a
wrong sign kinetic term. Alternatively, one can see it as coming from a perfect fluid source with negative energy density (see (\ref{enden}) below). We will restrict attention to  $\mu>1$ in what follows.

G\"{o}del's solution was originally obtained as a solution of gravity with negative cosmological constant $\L = -1/l^2$ in the presence
of a pressureless fluid source. It is instructive to check that the energy-momentum tensor of our scalar field solution $\T = x + i y$ behaves exactly as a pressureless fluid:
\bea
T_{\m\n} &\equiv &  - { 4 \p \over l_3 \sqrt{-g}}{\d S_\T \over \d g^{\m\n}} \nonu
&=& {4 \p (\m - 1)\over l_3 \T_2^2} \left[ \pa_{(\m}\T \pa_{\n)}\bar \T - \half g_{\m\n} \pa_\r \T \pa^\r \bar \T \right]
 \nonu
&=& \rho u_\m u_\n \, ,
\eea
where the unit vector is $u^\m = {2\over  l} \d^\m_0$ and the energy density of the fluid is
\be
\rho = { 16 \p (\m -1)\over \m l_3 l^2}\label{enden} \, .
\ee
Setting $\m = 2$ we again find the expression in \cite{Godel:1949ga}. The fluid flow is rotational since $ \star_3 ( u \wedge d u)$ is a nonzero constant,
indicating that  G\"{o}del space rotates around every point.

It is well-known that  G\"{o}del space suffers from causal pathologies in the form of closed timelike curves. These are most apparent in the coordinate
system which has the Poincar\'e disk as the  spatial base.
The following coordinate transformation takes us to this frame:
\bea%
t&\to& t+2 \m \arg(1 - z), \\
w &\to&i\frac{1+z}{1-z}\label{uhptodisk} .
\eea%
We define  $z= r e^{i \varphi}$ to get%
\be ds^2=\frac{ l^2}{4}\left[-(d
t+\mu \frac{2 r^2}{1-
r^2}d\varphi)^2 +4 \mu \frac{d r^2+
r^2d\varphi^2}{(1- r^2)^2}\right]\,.\label{Godelcoords} \ee

In the above form of the metric, it is easy to see that the vector field $\pa_{\varphi}$ becomes timelike for $r > \frac{1}{\sqrt{\m}}$, so that  $ \varphi$-circles become closed timelike
curves for these values of the radius.
The disk coordinate frame is also useful to visualize the axion-dilaton solution.
The brane source is located at the point $z = 1$ on the boundary. The lines of constant dilaton $\T_2$ are circles tangent to $z=1$. Using (\ref{Fvstau1}),(\ref{godelUHPmain}) one can show that
 $\T_2$ also plays the role of the scalar  potential for the electric field, so these circles are also equipotential surfaces. The electric field lines
are the lines of constant $\T_1$ and are orthogonal to the equipotential circles.
These properties are illustrated in \figref{fieldlines}.

\FIGURE{
\includegraphics[scale=0.5]{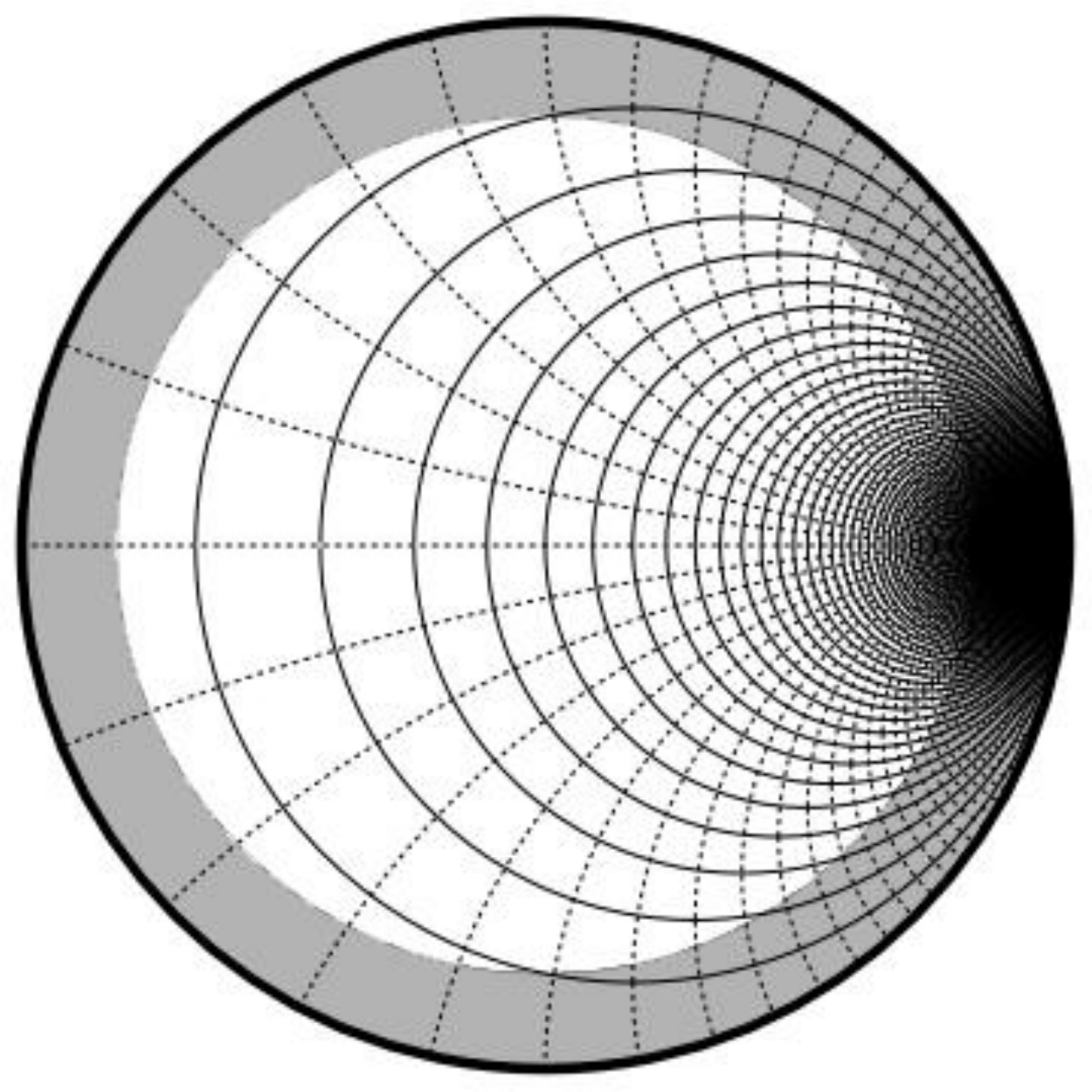}\caption{The  G\"{o}del solution in disk coordinates. Circles within the gray zone and centered at the origin
are closed timelike curves.  The solid lines are equipotential surfaces (constant $\T_2$), the dotted lines
are electric field lines (constant $\T_1$). The brane source is at $z=1$. }\label{fieldlines}}

\paragraph{The source:}
Since $\T$ has a pole at infinity in the upper half plane coordinates, we expect to have  a brane-like source of the form (\ref{sourceterms}) there. This was also nicely apparent in \figref{fieldlines}, where we plotted the electric fieldlines. Let us compute the total charge of this source.
It's convenient to  make an $SL(2,\Z )$ transformation and consider
\be
\T = - 1/w\label{godeltau} \, ,
\ee
so that the location of the pole is now at the origin. As $\tau_1$ is related to the $U(1)$ field sourced by M2-branes we expect the source to be made up of these. To be more precise, the eleven-dimensional definition of M2-brane charge is:
\be \label{chargeint}
q_{M2}=\int_{\Sigma_7}\star_{11} F_4\,.
\ee
Using the compactification ansatz \eqref{11Dreduction} we can rewrite this definition in three-dimensional form:
\be
q_{M2}=\int_{\g}  d\tau_1\,,
\ee
here $\gamma$ is a curve in the spatial base of the three-dimensional metric. Plugging in the explicit form (\ref{godeltau}) of $\tau_1$ for the G\"odel background and taking the limit where $\gamma$ becomes the real line one finds that
\be
q_{M2}=\int_{-\infty}^{\infty}\frac{dx}{x^2}=+\infty\,.
\ee
Hence we should see the pole in $\T$ as coming from a source of infinite charge.
The same conclusion is of course reached by computing the source for the dilaton field. The fact that the global G\"odel space carries an infinite amount of charge is probably due to the fact that the axion-dilaton solution covers an infinite number of fundamental SL$(2,\Z )$ domains. It seems plausible that one can obtain finite charge solutions by taking appropriate quotients of the global G\"odel space.

\section{Joining G\"odel to AdS} \label{sec-glue}
One of our original motivations in studying the system (\ref{3daction}) was to analyse solutions corresponding to branes wrapped around the S$^2$ of an AdS$_3\times $S$^2$ geometry.
Since such states carry a non-trivial topological charge that appears as a central element of the supersymmetry algebra it was to be expected that such solutions are no longer asymptotically AdS.
However, both from the point of view of holography and from the black hole microstate motivation it would be interesting if there was some kind of `embedding' of these solutions into an asymptotic
AdS spacetime. Probably the most straightforward way of realising such a setup is by enclosing a G\"odel region carrying the M2-charge by a domainwall that cancels this charge. Then, as in three
 dimensions all vacuum spacetimes are locally AdS$_3$, on the other side of the wall we are guaranteed to find a local AdS$_3$ spacetime. In this section we will realise exactly this idea, although it
 turns out that, under our assumptions, demanding that the AdS-side of the domainwall is connected to the boundary is equivalent to having a negative tension domainwall. In the case we have the G\"odel part of
 spacetime on the outside then the domainwall is made up of more familiar positive tension, smeared out M2-branes. For an overview see \figref{gluedd}.
 
\FIGURE{
\includegraphics[scale=0.8]{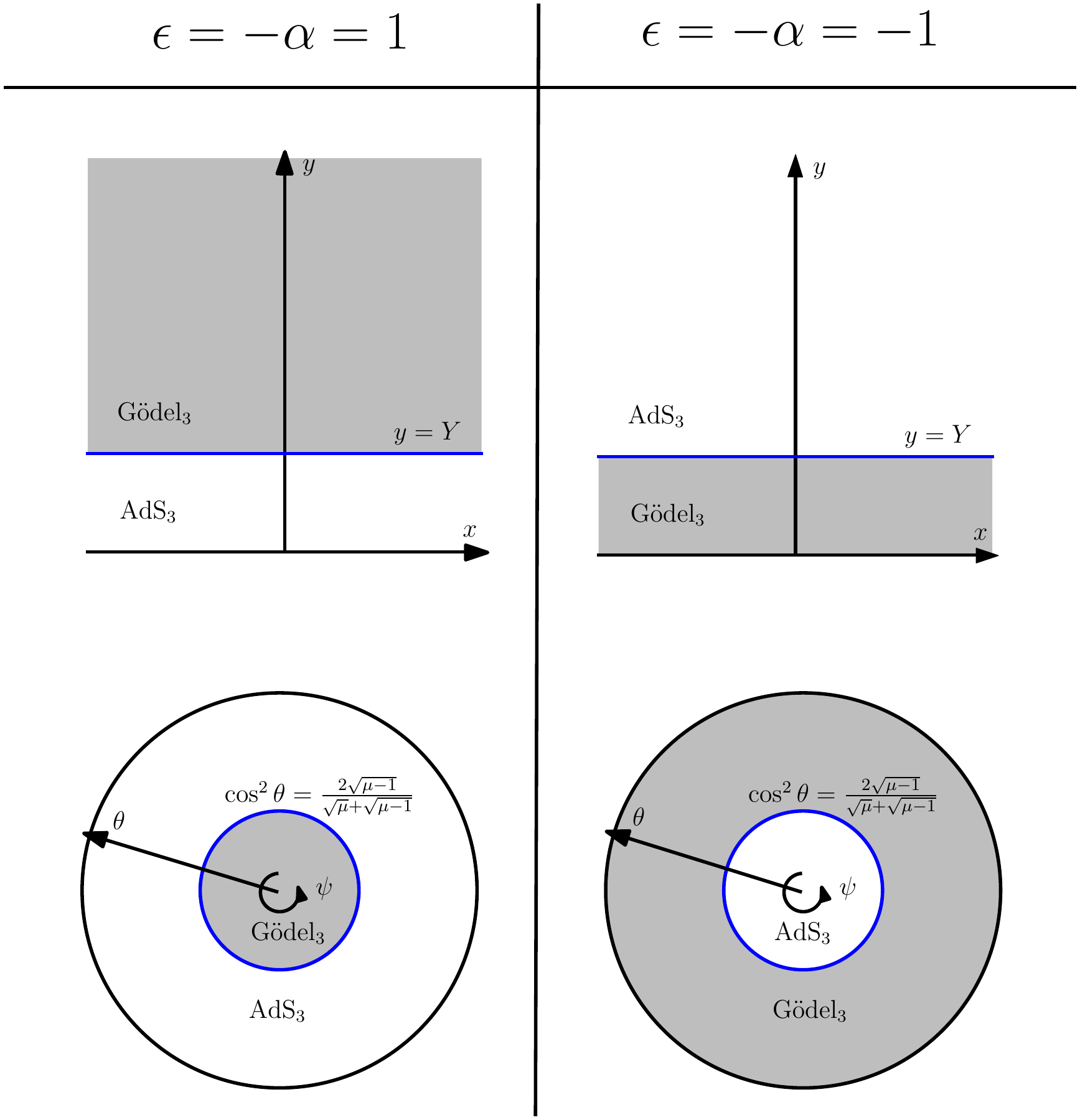}\caption{This figure gives an overview of the different gluings. On top we have presented the picture in the upper half plane coordinates $x, y$, most natural to G\"odel space at $t=$cst. As discussed in the main text, in these coordinates one still has to make an identification. On the bottom the glued spacetime is presented in global AdS coordinates. The disk shown here is a $\sigma=$cst slice of the cylinder that is AdS$_3$. Note that we have introduced the standard Penrose coordinate $\theta$, defined by $\tan\theta=\sinh \r$. The left hand side is that for the choice of $\epsilon=1$ and $\alpha=-1$. In this case the AdS part of space connects to the boundary, but the domainwall has negative tension. On the right hand side the situation is depicted for the opposite choice. Now the G\"odel part connects to the boundary and the domainwall has positive tension and an interpretation as smeared M2-particles.}\label{gluedd}}

Another motivation to consider such a domainwall construction is the analogy to \cite{Boyda:2002ba,Drukker:2003sc,Gimon:2004if}. In these references the authors show that one can remove the closed timelike
curves of G\"odel-like spacetimes by introducing a domainwall that connects it to an AdS-like spacetime. Naively one would hope the same effect to take place in the current setup. However, using the specific domainwall
ansatz below, this seems not to be the case, as closed timelike curves are present in the spacetime even after introducing the domainwall. The authors were not able to find a more general construction that eliminates
the closed timelike curves. This however remains an interesting goal for future research. It would also be interesting to see if there are smooth interpolating solutions in the spirit of \cite{Berglund:2005vb,Bena:2005va, Balasubramanian:2006gi},
describing these domainwall configurations.

\paragraph{The domainwall:}
More concretely, when trying to connect, or `glue', the G\"odel
space through a domainwall to AdS$_3$, it is clear that we need to
cancel M2-charge on one side of the wall, and hence the domainwall
should carry this charge. It thus seems natural to assume that the
domainwall is made up of M2-branes that wrap the internal S$^2$.
Since these branes are point particles in the three non-compact
dimensions we need to smear them along a spacelike direction to
make a domainwall\footnote{As the smearing does not induce tension
along that direction some might rather prefer to call such a
configuration a charged dust shell instead of a domainwall. } out
of them. By these assumptions we extend our previous Lagrangian
(\ref{3daction}) with a domainwall source made up of M2-particles
(\ref{sourceterms}): \bea
\frac{l_3 L}{2\pi}&=&\sqrt{-g}\left(R+\frac{2}{l^2}-(\mu-1)\frac{\partial_\m\tau_2\partial^\m\bar\tau_2}{\tau_2^2}-\frac{(\m-1)\tau_2^2}{2}F_{\m\n}F^{\m\n}\right)\label{domainwallaction}\\
&&-2\a(\mu-1)\int\,d\l\,d\s\, \delta^3(x^\m-X^\m(\s,\l))\left(\frac{\sqrt{-g_{\m\n}(X)\partial_\s X^\m \partial_\s X^\n}}{\tau_2(X)}-\b\frac{A_\m(X)}{2}\partial_\s X^\m\right)\nonumber \, .
\eea
Note that here we found it more convenient to replace the real scalar $\tau_1$ with its Hodge dual $F=dA$ as in (\ref{Fvstau1}), $\sigma$ parametrizes the M2-worldline and $\l$ is a smearing parameter.

For the moment we will leave the tension $\a$ arbitrary and take $\beta=\pm 1$. Furthermore, we introduced the parameter $\mu$ also in the domainwall action as three-dimensional solutions for abritary values of $\m$ might be of
independent interest. As before, the main value of interest is $\mu=3/2$, which is the only value that corresponds to the M-theory embedding as given in section \ref{secto3d}. In that case (and when $\a>0$), the domainwall can be
 interpreted as smeared M2-branes (or anti M2 branes depending on the sign of $\b$).

Since the domainwall source couples to the dilaton $\tau_2$, which in the G\"odel solution is given by $\tau_2=y$, one can only smear with a constant density along a surface of constant $y$. For the rest of this section we
will be working in the specific coordinates $(x^0,x^1,x^2)=(t,x,y)$, known from the G\"odel solution (\ref{godelUHPmain}), and assume the following embedding of the domainwall:
\be
X^0(\s,\l)=\s\,,\qquad X^1(\s,\l)=\l\,,\qquad X^2(\s,\l)=Y\label{scalans} \, .
\ee
Furthermore, we will assume all fields to be continuous in $y$ in spacetime, although we will allow derivatives of the fields to be discontinuous at $y=Y$. This can be achieved by `gluing' two different solutions of the
Lagrangian \eqref{3daction} (\eqref{domainwallaction}) without domainwall source) together at $y=Y$. More concretely we propose the following field configuration:
\bea
ds^2&=&N^2dy^2+h_{ab}dx^adx^b\nonumber ,\\
N&=&N_{\mbox{\tiny G\"odel}}\Theta[\epsilon\,(y-Y)]+N_\mathrm{AdS}\Theta[\epsilon\,(Y-y)]\nonumber , \\ h_{ab}&=&h_{ab}^{\mbox{\tiny G\"odel}}\Theta[\epsilon\,(y-Y)]+h_{ab}^\mathrm{AdS}\Theta[\epsilon\,(Y-y)]\nonumber , \\
\label{glueansatz}\\
A&=&\left(\frac{l}{y}\,\Theta[\epsilon(y-Y)]+\frac{l}{Y}\,\Theta[\epsilon(Y-y)]\right)dt+\left(\frac{l\m}{2y^2}\,\Theta[\epsilon(y-Y)]+\frac{l\m}{2Y^2}\,\Theta[\epsilon(Y-y)]\right)dx\nonumber ,\\
\tau_2&=&y\,\Theta[\epsilon(y-Y)]+Y\,\Theta[\epsilon(Y-y)]\nonumber \, .
\eea
Here the Heaviside stepfunction, $\Theta(x)$, gives a formal way of defining the gluing. We introduced a sign $\epsilon=\pm1$ to distinguish between the solution with AdS/G\"odel `inside' or `outside'.
I.e. when $\epsilon=1$ we have G\"odel at $y>Y$ (inside) and AdS at $y<Y$ (outside), when $\epsilon=-1$ the situation is exactly opposite. Note that on the AdS side, we take the complex scalar $\tau$ (written here as $\tau_2$ and a gauge field $A$) to be constant. As we will show below, the chosen values of these fields make sure the ansatz for $A$ and $\tau_2$ is valid and satisfies the corresponding equations of motion. Note that in the above ansatz the only remaining unknowns are $N_\mathrm{AdS}$ and $h_{ab}^\mathrm{AdS}$. Their G\"odel counterparts are known by definition and can
be read off from (\ref{godelUHPmain}). Although we know on general grounds that the metric on the other side will be AdS, we do not know the precise form of the functions $N$ and $h_{ab}$ on that side, as $t,x$ and $y$
are an unknown set of coordinates there. As we will explicitly work out below, the precise form of these functions can be found by solving the Einstein equations with suitable boundary conditions. Of course, by definition of our gluing procedure, we want the total metric to be continuous, so this imposes the condition:
\bea
&&N_\mathrm{AdS}|_{y=Y}=N_{\mathrm{Godel}}|_{y=Y}=\frac{l^2}{4Y^2}\label{boundcont} \, ,\\
&&h_{ab}^\mathrm{AdS}dx^adx^b|_{y=Y}=h_{ab}^\mathrm{Godel}dx^adx^b|_{y=Y}=-\frac{l^2}{4}\left(dt^2+\frac{2 \m dxdt}{Y}+\frac{\m(\m-1)}{Y^2}dx^2\right)\nonumber \, .
\eea

\paragraph{The junction conditions:}
We analyse the equations of motion following from (\ref{domainwallaction}) for the ansatz (\ref{glueansatz}). Each of these equations can be split into  three separate ones that have to be obeyed in the different regions:
two bulk equations evaluated at $y>Y$ and $y<Y$ respectively and one `singular' equation at $y=Y$. The bulk equations are the equations of motion  following
from the Lagrangian (\ref{3daction}) without the domainwall source. Our ansatz automatically satisfies the bulk equation in the G\"{o}del region, while in the AdS region
$N_\mathrm{AdS}$ and $h_{ab}^\mathrm{AdS}$ have to be chosen so as to satisfy the vacuum Einstein equations. The singular equation at $y = Y$  is proportional to $\delta(y-Y)$ and receives two contributions: the first comes from the domainwall term in the Lagrangian, the second from possible singular derivatives of the fields at $y=Y$. In general relativity these singular equations that impose the equality between the domainwall source terms and the discontinuities of field derivatives are known as the Israel junction conditions \cite{Israel:1966rt}. In \appref{IsJuncAp} we review the derivation of these equations and extend them to more general field theories. Note that the ansatz for the gauge field $A$ and scalar $\tau_2$ in (\ref{glueansatz}) was chosen in such a way that continuity across $y=Y$ is assured.

As mentioned above, only the metric is not completely determined yet in our ansatz. Before deriving the correct form of the metric by the Israel junction conditions let us first check that the ans\"atze for the other fields satisfy their corresponding junction conditions. Applying the general formula (\ref{GenJuncCond}) to the Lagrangian (\ref{domainwallaction}) and the fields $\tau_2$ and $A$ we get the following two equations (all evaluated at $y=Y$):
\bea
\frac{\Delta \partial_{y}\tau_2}{N}&=&-\a\frac{\sqrt{-h_{00}}}{\sqrt{-h}}\\
N\sqrt{-h}\tau_2^2\Delta F^{y\m}&=&-\a\b\,\delta^{\m}_t.
\eea
The definition of the operator $\Delta$ can be found in (\ref{defdel}). Plugging in our gluing ansatz (\ref{glueansatz}) these become
\bea
\epsilon&=&-\a\label{epsa}\\
\b&=&1.
\eea
So we see that by choosing the correct tension and charge the ansatz can be a solution for both signs of $\epsilon$. Note however, that demanding the part of spacetime connected to the boundary at $y=0$ to be AdS, corresponds to $\epsilon=1=-\a$ and hence implies that the domainwall is made up of particles with negative tension.

Another equation that needs to be checked is that for the embedding scalars $X^\m$. Based on the assumption of constant smearing density we made the ansatz (\ref{scalans}) for these scalars,
but we still have to verify if this is consistent with the equations of motion obtained by varying the Lagrangian (\ref{domainwallaction}) with respect to the $X^\m$. This equation can be seen as a special case of a junction equation, but with only a contribution from the domainwall and no singular bulk terms. Plugging the ansatz (\ref{scalans}) and the values of the other fields into those equations of motion leads directly to the condition
\be
\b=1 .
\ee
This is nicely consistent with the value of $\beta$ found above.

What remains is to solve the bulk Einstein equations on the AdS side with boundary conditions specified by the Israel matching conditions. Applying the Israel junction conditions (\ref{IsJuncCond}) to the case at hand we find
\be
\Delta(K^{ab}-g^{ab}K)=\frac{-\a }{2 \tau_2 N \sqrt{-h_{00}}}\delta^{a}_{t}\delta^{b}_t\label{match} \, .
\ee
These equations together with (\ref{boundcont}) should be interpreted to provide boundary conditions at $y=Y$ for the local AdS metric appearing in our ansatz (\ref{glueansatz}).

\paragraph{The G\"odel-AdS solution:}
Solving the vacuum Einstein equations with the boundary conditions (\ref{match}) and (\ref{boundcont}) at $y=Y$ is rather straightforward. One can verify that the solution for $\e (Y - y) >0$  is given by
\be
ds^2=\frac{l^2}{4}(-(dt+\m\frac{dx}{y})^2+\frac{\m}{y^2}(f(y)dx^2+f^{-1}(y)dy^2)) \, ,
\ee
with
\be
f(y)=\m  +(1-\mu)\frac{y^2}{Y^2} \, .
\ee
Although not the most familiar form, one can check that this metric is locally AdS$_3$ by calculating the Ricci tensor. Furthermore it can be put in the standard global AdS coordinates \eqref{globalmetr} by a simple coordinate transformation. Define the quantity $\o=\frac{\sqrt{\m(\m-1)}}{Y}$. The transformation is then given by
\bea
\cosh 2\r&=&\frac{\m}{\o y} , \\
\sigma&=&\frac{\o x+t}{2} ,\\
\psi&=&\frac{\o x-t}{2} .
\eea

This coordinate transformation shows that the domainwall at $y=Y$ maps to a cylindrical hypersurface at constant radius $\rho_\star$ in global AdS$_3$, with $\cosh2\rho_\star=\sqrt{\frac{\m}{\m-1}}$. Note that in these coordinates the position of the domainwall is independent of the choice of $Y$. This is due to the fact that one could have removed $Y$ from the previous discussion by the transformation $x\rightarrow Yx$, $y\rightarrow Yy$, which is a symmetry of G\"odel-space. We
give an overview of the different gluings in the different coordinate systems in figure \figref{gluedd}

As one knows from the standard coordinates on AdS$_3$, the coordinate $\psi$ is periodic: $\psi\sim\psi+2\pi\,\n$. The most familiar value is $\nu=1$, which corresponds to global AdS.
However one can also choose $\n<1$, in which case the geometry corresponds to a conical defect geometry.
Continuity requires that we impose the same periodicity on G\"{o}del side as well; this leads to
the identification
\bea
t &\sim& t - 2 \p \n ,\nonu
x &\sim& x + {2 \p \over \o}\n \label{godelident}.
\eea
It is however straightforward to check that this identification leads to new CTCs
in the G\"odel part of the glued spacetime. This holds in both cases of gluing, the one with G\"odel at small $y$ ($\epsilon=-1$) and that at large $y$ ($\epsilon=1$). The reason is, that the Killing
vector $\frac{1}{\omega}\partial_{x}-\partial_t$, that generates the identification, is timelike in G\"odel both for $y\rightarrow 0$ and $y\rightarrow\infty$.\footnote{Note that one could generalise the
identification in AdS to $(\psi,\sigma)\sim(\psi+2\pi\n_1,\s+2\pi\n_2)$. The metric would in this case become a spinning conical geometry, but it is easy to check that such an identification leads in general to
closed timelike curves in both the G\"odel part and the AdS part of the glued spacetime.}

Let us also discuss the preservation of supersymmetry  in our glued
solutions. Since we need to impose the extra identification (\ref{godelident}) on the G\"{o}del part,
Killing spinors should be well-defined (i.e. either periodic or antiperiodic) under this identification.
In the case $\nu = 1$, we have global AdS$_3$ on one side, preserving 8 supersymmetries. On the G\"{o}del side,
all  4 Killing spinors (\ref{probeKSmain},\ref{KSmain}) are antiperiodic under the identifications and are therefore globally well-defined.
When $\n <1$, the  Killing spinors in the AdS$_3$ part will pick up a complex phase under transport around the conical defect.
 As is well-known \cite{David:1999zb}, in order to have globally well-defined Killing spinors   one must compensate for this by turning on a Wilson line for one
of the three-dimensional gauge fields that arise from the reduction of the five-dimensional metric on the two-sphere. From the five-dimensional point of view, this is nothing but
a `large' diffeomorphism of the form $\varphi \to \varphi \pm t$, where $\varphi$ is the angular coordinate on the two sphere. After performing this diffeomorphism,
one can see that 4 out of 8 local supersymmetries exist globally on the AdS side. Doing the same diffeomorphism on the G\"{o}del side
preserves 2 out of 4 local Killing spinors, as one can see from (\ref{KSmain}). So we conclude that our glued geometries are supersymmetric, with the G\"{o}del
part preserving half of the supersymmetries of the AdS part in each case.

The case we would be most interested in is that of positive $\epsilon$, as then the glued spacetime is asymptotically AdS and we know how to do holography on such spaces.
However, as can be read from the condition (\ref{epsa}) and the Lagrangian (\ref{domainwallaction}), $\epsilon=1$ implies negative tension for the domainwall. Even though negative tension domain walls are not unheard of, either in supersymmetric theories (see e.g. \cite{Shifman:1999ri}), or as orientifold-type objects in string/M theory (see e.g. \cite{Dai:1989ua,Horava:1995qa}),
clearly it is harder to interpret them in terms of fundamental M-theory branes.
It might still be interesting to understand these glued spaces in more detail through holography.
 Also, there is a potential danger of instabilities as discussed in \cite{Marolf:2001ne}, however since our construction preserves supersymmetry
on both sides of the wall we believe this is not an issue in our case.

\section{Discussion and future directions} \label{sec-con}

In this work, we have constructed supersymmetric solutions to three-dimensional axion-dilaton gravity with negative cosmological constant
which describe the backreaction of S$^2$-wrapped M2-branes in M-theory. We found a class of solutions where the axion-dilaton is (anti-) holomorphic
and where the local geometry is that of the three-dimensional G\"{o}del universe.
We showed that these solutions preserve four supersymmetries in agreement with the analysis in the probe approximation.
 We have also shown that our solutions can be glued, in a supersymmetric manner,
into asymptotically AdS$_3$ geometries by including a charged domainwall.

Let us comment on some aspects which deserve a better understanding and some interesting directions for future research. A first puzzle
is that our backreacted solutions have M2-brane sources only on the boundary, whereas in the probe approximation discussed in \secref{sec-probe}, it appeared as if the M2-branes could
be placed anywhere.
This could point to the existence of more general solutions with sources in the interior, but it could also be due to the fact that these are codimension-two objects producing long-range fields;
hence the probe picture might be unreliable.
Another feature of our solutions is that the brane charge residing on the boundary is actually infinite. This can be seen as a consequence of the fact that $\T$ takes values in the entire upper half
plane. One way to obtain a `stringy' finite charge solution, would be to identify values of $\T$ related by $SL(2,\Z )$ transformations and make a similar identification on the coordinate $w$ of the base manifold.
The charge integral \eqref{chargeint} would then be finite. A similar procedure would work for an arithmetic subgroup of $SL(2,\Z )$. One should note that such constructions in general involve identifications generated by
timelike vectors and will produce  more closed timelike curves.
Nevertheless, such configurations are finite-energy, finite charge BPS solutions, and one would expect them to contribute to the path integral.
 It would be interesting to  understand their role  better.

Our original motivation for studying this system was the black hole microstate or deconstruction proposal \cite{Gaiotto:2004pc,Gaiotto:2004ij,Denef:2007yt,Gimon:2007mha}, where it was argued that S$^2$ wrapped M2-brane probes have a large quantum mechanical degeneracy (coming
from lowest Landau levels on the internal Calabi-Yau) that can account for the black hole entropy. An interesting question is whether this degeneracy can also be
understood after including the backreaction. This might furthermore clarify the relation between these deconstruction states and other BPS solutions carrying the black hole charges that are more closely related to the original fuzzball proposal \cite{Mathur:2005zp,
Balasubramanian:2008da,Mathur:2008nj}. Although various BPS solutions were explicitly constructed, see e.g.
\cite{Lunin:2001jy,Giusto:2004kj,Giusto:2004ip, Giusto:2004id,Bena:2004de,Berglund:2005vb,Bena:2005va,Balasubramanian:2006gi,Rychkov:2005ji}, it was argued recently that these might only account for a subleading fraction of the black hole ensemble \cite{deBoer:2008zn,deBoer:2009un}.
It would in particular be interesting to understand the deconstruction microstates in a dual CFT and see if and how they evade the bound of \cite{deBoer:2009un}.
Our gluing procedure in \secref{sec-glue} should be seen as
a first attempt in this direction.  It would be interesting to explore more general gluings into AdS$_3$ and the holographic interpretation of these constructions.
Another approach to understand the quantum properties of the system (\ref{3daction})  would be to argue the existence of a  dual CFT  in the spirit of
the Kerr-CFT correspondence \cite{Guica:2008mu}. It is expected that, under suitable boundary
conditions, the $U(1)_L \times SL(2,\R )_R$ symmetry of the G\"{o}del  solution will extend to an asymptotic Virasoro algebra
(see \cite{Compere:2007in} for an example involving a G\"{o}del solution
with a different matter content). Since our system is supersymmetric for $\m = 3/2$, one would  expect the asymptotic  algebra to extend to a superconformal
algebra in that case. It would also be interesting to see if the theory has interesting excitations, such as black hole solutions, which could be understood in the dual CFT.

\section*{Acknowledgements}
The authors are especially grateful to E.~Gimon and T.~Wyder for collaboration on part of this work.
During this work we benefitted from many discussions. It is a pleasure to thank D.~Anninos, G.~Compere, R.~de~Mello~Koch, F.~Denef, S.~Detournay, S.~El-Showk, T.~Erler, B.~Gaasbeek, M.~Guica, M.~Kleban, J.~Manschot, G.~Moore, P.~Smyth and A.~Van Proeyen.
TSL was supported in part by the Natural Sciences and Engineering Research Council of Canada and the Institute of
Particle Physics.
JR was supported under a grant of the National Research Foundation of South Africa. His research has been supported in part by the EURYI grant EYI/07/E010 from EUROHORC and ESF.
The work of DVdB is supported by the DOE under grant DE-FG02-96ER40949.
The work of WVH and BV is supported in part by the FWO - Vlaanderen, project G.0235.05 and in part by the Federal Office for Scientific, Technical and Cultural Affairs through the 'Interuniversity Attraction Poles Programme – Belgian Science Policy' P6/11-P. B.V. is Aspirant of the FWO Vlaanderen and thanks the High Energy Theory Group at Harvard University for its hospitality.

\begin{appendix}
\section{Other embeddings of the 3D system}\label{app-duality}
The three-dimensional theory (\ref{3daction}) can arise as a
consistent truncation of
 other theories than the M-theory context discussed in the main text. Here, we will
discuss a few interesting alternative embeddings in type IIB
string theory compactified on S$^3 \times \calm_4$ where $\calm_4$
is a compact 4-manifold. Such backgrounds arise for example in the
near-horizon limit of the  D1-D5 system and are of relevance in
the description of the five-dimensional Strominger-Vafa  black
hole \cite{Strominger:1996sh}. These alternative embeddings will
be derived by applying U-dualities and the 4D-5D connection (see
\cite{Raeymaekers:2007ga,Raeymaekers:2008gk} for more details on
this procedure). We will first discuss the effect of these
transformations on the `background branes' that have an AdS$_3$
near-horizon region, and then discuss the transformation of the
`source' branes that couple to the three-dimensional axion-dilaton
and whose backreaction turns AdS$_3$ into G\"{o}del space.

We start with our M-theory embedding, replacing the Calabi-Yau
manifold by $ S_M \times S_5 \times\calm_4$, where $S_M$ and $S_5$
are circles. For simplicity we take $\calm_4$ to be a four torus,
but most of the discussion applies to the case of $K_3$ as well.
In particular, we take $\calm_4 = T_2 \times T_3$ where $T_2, \
T_3$ are two-tori.  For the background branes, we take M5-branes
wrapping internal 4-cycles, extending along a common direction to
form a BPS black string in five dimensions. For simplicity, we
restrict to three types of M5-brane charges: we take $p^1$
M5-branes wrapping $T_2 \times T_3$ and  $p^a, \ a = 2,3$
M5-branes wrapping $ S_M \times S_5 \times T_a$. The near-horizon
geometry  is an  AdS$_3 \times $S$^2  \times S_M \times S_5 \times
\calm_4$ background of the form (\ref{11Dreduction}) with constant
$\T$.

We perform the  limits and dualities summarized in
\tableref{Udual}: First, we compactify on $S_M$ to type IIA: $p^1$ are now
NS5-branes and the $p^a$ become D4-branes. Then we T-dualize on
$S_5$ to get type IIB  on the dual circle $\tilde S_5$.
 $p^1$ becomes a KK monopole charge and
    $p^a$ are now D3-branes. Because of the presence of KK monopole charge, we can apply the standard 4D-5D connection \cite{Gaiotto:2005gf} and obtain a five-dimensional configuration by taking the asymptotic size of $\tilde S_5$  to infinity.
 We then have a configuration of intersecting D3-branes forming a black string
 in a six-dimensional compactification of type IIB
on $ \calm_4$. The near-horizon geometry is AdS$_3 \times $S$^3/p^1
\times \calm_4$. By taking $p^1 =1$, we get precisely AdS$_3
\times $S$^3 \times \calm_4$.

Let us follow what happens to the source branes under the above
series of dualities. Starting from an
 M2-brane wrapping the S$^2$ in the M-theory frame, one obtains a D3-brane wrapping the S$^3$ in the type IIB frame. Hence in this frame, all the branes in the system are D3's!
Starting from this configuration, one obtains some useful
embeddings by applying further dualities. The results are
summarized in \tableref{interestingframes} in the main text.

  First, let's do two further T-dualities on $T_2$. The background
branes are now the well-known D1-D5 system. The source branes have
become D5 branes wrapping S$^3 \times T_2$. In the probe picture,
such branes were argued  to have a high lowest Landau level
degeneracy and account for the entropy of the five-dimensional
Strominger-Vafa black hole \cite{Raeymaekers:2007ga}. If we
perform a further S-duality, the background consists of
fundamental strings and NS5-branes, while the source branes are
NS5-branes on S$^3 \times T_2$, coupling to the complexified
K\"{a}hler modulus of $T_3$. This frame is interesting because all
the excitations are in the NS-sector, allowing for a sigma-model
description of the system.

One can also obtain an F-theory embedding by starting from the
intersecting D3-brane description and doing four T-dualities along
$T_2 \times T_3$. This just interchanges the two types of
background D3-branes, while the source branes now become
D7-branes wrapping S$^3 \times \calm_4$. In this frame,   $\T$ is
the standard axion-dilaton of type
 IIB, and the resulting solutions can be seen as nontrivial F-theory backgrounds involving D7-branes. It would be interesting to
study the geometry of the fibration of the F-theory torus  over
G\"{o}del space described by the $\T$ field in more detail.

{\small
\begin{table}
\begin{center}
\begin{tabular}{lclclcl}
M/$  S_M \times S_5 \times \calm_4$& $\longrightarrow$& IIA /$ S_5
\times \calm_4 $&$\longrightarrow$ & IIB/$  \tilde S_5
\times\calm_4$ &
$\longrightarrow$ &IIB/$ \calm_4$ \\
&$ S_M \to 0$ & &T($S_5$) & &$\tilde S_5\to \infty $ & \\
$p^1$: M5 /$\calm_4$ & $\longrightarrow$& NS5  /$ \calm_4$  & $\longrightarrow$&   KK5  /$ \calm_4$  &$\longrightarrow$ &  (def. angle)\\
$p^a$:M5 /$ S_M \times S_5 \times T_a$  &  & D4 /$S_5 \times T_a$&  &  D3 /$ T_a$& &  D3/$ T_a$\\
\end{tabular}
\end{center}\caption{Dualities and 4D-5D connection}\label{Udual}\end{table}
}

\section{Details of the supersymmetry analysis} \label{app-susy}
In this appendix we discuss the supersymmetry properties of our
solutions. We will mostly use the eleven-dimensional supergravity
point of view and comment on the analysis in five-dimensional $\caln
= 1$ supergravity in \appref{app-5Dsusy}.
\subsection*{Conventions}
We will work in the eleven-dimensional supergravity  conventions of
\cite{Gauntlett:2002fz}. The bosonic part of the eleven-dimensional
supergravity (M-theory) action was given in (\ref{Maction}). The
Killing spinor equation is \be \nabla_M \e + {l_M^3 \over
12}\left[ \G_M\sF{}_4 - 3\sF{}_M \right] \e = 0 \label{susyvar} \,
,\ee where $\sF{} = { 1 \over 4 !} F_{MNPQ}\G^{MNPQ}, \sF{}_M = {
1 \over 3 !} F_{MNPQ}\G^{NPQ}$ and $\e$ is an eleven-dimensional spinor
satisfying the Majorana condition \be \e^* = B \e\label{Maj11D} \,
, \ee with $B$ a matrix satisfying $B \G^{ M} B^{-1} =  \G^{ M *}$
which we will specify later.

We will consider bosonic field configurations of the form
(\ref{11Dreduction}) with a three-dimensional metric of the form
(\ref{metricansatz}). To further simplify expressions, we will
also use the explicit form of our solutions
(\ref{tausol}),(\ref{chisol}),(\ref{phisol}). We will show that
all such configurations are supersymmetric solutions of eleven-dimensional
supergravity. When $\T$ is constant, the geometry is locally
AdS$_3 \times $S$^2 \times CY_3$ and the solution preserves 8 local
Killing spinors. When $\T$ is not constant, AdS$_3$ is replaced
by a local G\"{o}del geometry and the number of local  Killing spinors is 4. 

It will be useful to work in a spinor basis adapted to the
decomposition  $SO(1,10)\supset SO(1,4) \times SO(6)$. We
decompose the eleven-dimensional Clifford matrices as (hatted indices refer to an
orthonormal frame) \bea
\G^{\hat \m} &=& \g^{\hat \m} \otimes \g_{(6)} , \qquad \hat \m = 0,\ldots,4 \, ,\\
\G^{\hat \a}&=& 1 \otimes \g^{\hat a} , \qquad \hat a = 5, \ldots,
10 \, , \eea where the $\g^{\hat \m}$ are $4 \times 4$ matrices
generating an $SO(1,4)$ Clifford algebra,  $\g^{\hat a}$ are $8
\times 8$ matrices generating an $SO(6)$ Clifford algebra and $
\g_{(6)} = i \g^{\hat 5 \ldots \hat{10}}$.

In what follows, we will take a basis where  $\g^{\hat 0},
\g^{\hat 1}, \g^{\hat 2}, \g^{\hat 3}$  are real (i.e. the usual
four-dimensional Majorana representation). For $\g^{\hat 4}$ we take $\g^{\hat
4} = i \g^{\hat 0 \hat1 \hat 2 \hat3}$, hence it is imaginary. The
matrix $B$ will be of the form \be B = \g^{\hat 4} \otimes
B_{(6)}. \label{Bmatrix} \ee

The Calabi-Yau manifold on which we compactify has two covariantly
constant $SO(6)$ spinors $\eta_+, \eta_-$. These can be taken to
satisfy the following properties that will be useful later \be
\begin{array}{lcl}
\nabla_a \eta_\pm = \nabla_{\bar a} \eta_\pm = 0 &\qquad& \g_{(6)} \eta_\pm = \pm \eta_\pm \\
\g_{\bar a}\eta_+ = \g_{ a}\eta_- =0 &\qquad& B_{(6)} \eta_\pm = \mp i  \eta_\mp \\
(\eta_\pm)^* = \eta_\pm &\qquad& \slashed J{} \eta_\pm = \pm {3 i
\over  l_M^2 \T_2^{1/3}}\eta_\pm.
\end{array}
 \label{etaprop}
\ee

We make the following ansatz for our eleven-dimensional Killing spinors \be \e =
\varepsilon^1 \otimes \eta_+ + \varepsilon^2 \otimes \eta_-
\label{spinordecomp}\,, \ee where $\varepsilon^i$ are spinors of
the Clifford algebra in five dimensions. Using doublet notation
for the five-dimensional spinors \be \varepsilon = \left( \begin{array}{c}
\varepsilon^1 \\ \varepsilon^2 \end{array}\right) \ee
 the eleven-dimensional Majorana condition (\ref{Bmatrix})  reduces to a symplectic Majorana condition for the five-dimensional spinors
\be \ve^*=  \g^{ \hat 4} \otimes \s_2 \ve \label{Maj5d} \, . \ee

We discuss the Killing spinor equations
 for the vacuum case (i.e. constant $\tau$), and afterwards we proceed to non-constant $\tau$ thus showing that the three-dimensional G\"odel space we obtained is supersymmetric.

\subsection*{The $\T=$ constant case: the AdS$_3 \times $S$^2$ vacuum}
\subsubsection*{Killing spinors}
We first consider the case where $\T$ is constant, when the
geometry is locally AdS$_3 \times $S$^2$. First we check the Killing
spinor equations with the index in the Calabi-Yau space. From our
decomposition (\ref{spinordecomp}) and the properties
(\ref{etaprop}) we find that these  equations are automatically
satisfied. Next we look at the Killing spinor equations with an
index in the five-dimensional space. With the choice of vielbein (hatted indices
are frame indices) \be
\begin{array}{lcl}
e^{\hat t} = {l \over 2 \T_2^{1/3}} ( dt + \chi ) ,&\qquad& e^{\hat \t} = {l \over 2 \T_2^{1/3}} d\t ,\\
e^{\hat z} =  {l \over 2 \T_2^{1/3}} e^{\f} dz , &\qquad & e^{\hat \varphi} = {l \over 2 \T_2^{1/3}} \sin \t d\varphi , \\
e^{\hat {\bar z}} =  {l \over 2 \T_2^{1/3}} e^{\f } d\bar z \,
,\nonu  &\qquad&
\end{array}
\ee (note the difference between the field $\f$ and the angular
coordinate $\varphi$) one finds, using (\ref{chieqn}), that the
spin connection 1-forms are \be
\begin{array}{lcl}
\o^{\hat t \hat z} = - {i \over 2} e^{\f } dz , &\qquad&  \o^{\hat \t \hat \varphi} = - \cos \t d \varphi ,\\
\o^{\hat t \hat{\bar z}} =  {i \over 2} e^{\f } d\bar z , &\qquad& \\
\o^{\hat z \hat{\bar z}} = -i( dt + \chi) + 2 (\pa \f dz - \bar
\pa \f d\bar z)  \, . &\ &
\end{array}\label{spinconn}
\ee We find that the  Killing spinor equations reduce to the
following conditions on the five-dimensional spinor doublet $\ve$: \be
\begin{array}{lcl}
\left(\pa_t - i \g_{\hat z \hat{\bar z}} \right)\ve = 0 ,&\qquad& \pa_\t \left( e^{- {i \t \over 2}  \g^{\hat \varphi}}\right)\ve = 0 ,\\
\left(\pa_z + (\pa \f - i \chi_z )\g_{\hat z \hat{\bar z}}
\right)\ve = 0 ,&\qquad&
\pa_\varphi \left(  e^{- {\varphi \over 2}  \g^{\hat \t \hat \varphi}} e^{- {i \t \over 2}  \g^{\hat \varphi}}\right)\ve = 0 , \\
\left(\pa_{\bar z} - (\bar \pa \f + i \chi_{\bar z} )\g_{\hat z
\hat{\bar z}} \right)\ve = 0.&\qquad&
\end{array}
\ee Using the solution  (\ref{chisol}) for $\chi$, these are
solved to give \be \varepsilon ={\sqrt{l} \over \sqrt{2}
\T_2^{1/6}} e^{i (t  + f ) \g_{\hat z \hat{\bar z}}} R(\t,
\varphi) \varepsilon_0 \, ,\label{5DKspinor} \ee where \be R(\t,
\varphi) =  e^{ {i \t \over 2}  \g^{\hat \varphi}}  e^{ {\varphi
\over 2}  \g^{\hat \t \hat \varphi}} \, , \ee and $\e_0$ is a
constant five-dimensional spinor doublet. Note that the ambiguity $\chi \to \chi
+ d f$ induces a shift in the phase of the Killing spinors. The
condition (\ref{Maj5d}) imposes that $\varepsilon_0$ satisfies the
symplectic Majorana condition \be \ve^*_0=  \g^{ \hat 4} \otimes
\s_2 \ve_0 \label{Maj5dKspin}. \ee Hence we find 8 independent
real Killing spinors. We see that our Killing spinors are
antiperiodic when $z$  goes around a circle, indicating that we
are in the Neveu-Schwarz sector, at least if we choose $f$ to have
vanishing monodromy. This is the case in the coordinate systems we
consider: both in the disc and upper half plane coordinate systems
(\ref{diskmain},\ref{UHPmain}), the function $f$ is zero.

\subsection*{The $SU(1,1|2)_L$ algebra}
The solutions with $\T$ constant have local Killing vectors
obeying the algebra of $SL(2,\R )_L \times SL(2,\R )_R \times
SU(2)_L$. The $SL(2,\R )_L \times SU(2)_L$ generators combine with
the 8 Killing spinors derived above to form the the supergroup
$SU(1,1|2)_L$.

The  $SL(2,\R )_L \times SU(2)_L$ Killing vectors are \be
\begin{array}{lcl}
l_0 = i \pa_t  , &\qquad& j_3 = i \pa_\varphi ,\\
l_+ = 2 e^{- \f} e^{i(t + f)}(\chi_z \pa_t - \pa_z ), &\qquad& j_\pm = e^{\mp i \varphi} (i \pa_\t \pm \cot \t \pa_\varphi ) ,\\
l_- = -2  e^{- \f} e^{-i(t + f)}(\chi_{\bar z} \pa_t - \pa_{\bar
z} ), &\qquad&
\end{array} \label{KV}
\ee where $f$ is the arbitrary real function entering in the
solution (\ref{chisol}) for $\chi$.

We introduce a convenient basis for our Killing spinors
(\ref{5DKspinor}) labelled by three doublet indices $m, \a, a$
taking values $\pm 1$: \be g_{m \over 2}^{\a a} = {\sqrt{l} \over
\sqrt{2} \T_2^{1/6}} \; e^{ {i m\over 2}( t + f)} e^{ - {i \a
\varphi\over 2} }  e^{ {i \t \over 2}  \g^{\hat \varphi}} g_{0\;
{m \over 2}}^{\; \a a} \, ,\label{Killspinorbasis} \ee where the
$g_{0\; {m \over 2}}^{\a a}$ are constant spinors satisfying \bea
\g_{\hat z \hat{\bar z}} g_{0\; {m \over 2}}^{\; \a a}&=&{m \over
2} g_{0\; {m \over 2}}^{\; \a a} , \nonu \g^{\hat \t \hat \varphi}
g_{0\; {m \over 2}}^{\; \a a}&=&- i \a g_{0\; {m \over 2}}^{\; \a
a}, \nonu \g_{(6)} g_{0\; {m \over 2}}^{\; \a a}&=&a g_{0\; {m
\over 2}}^{\; \a a}. \eea In our explicit spinor basis, we take
them to be \bea
 g_{0\; {m \over 2}}^{\; + a} &=& {1 \over \sqrt{2}} ( 0\ i\ m\ 0 )^T \otimes \eta_a ,\nonu
g_{0\; {m \over 2}}^{\; - a} &=& {1 \over \sqrt{2}} ( i\ 0\ 0 \
m)^T \otimes \eta_a.\label{constspinors} \eea

From the Killing vectors and Killing spinors one can compute the
isometry supergroup of the supergravity background using the
method developed in \cite{Gauntlett:1998kc}. In particular, we are
interested in the fermionic anticommutators, which can be computed
by evaluating spinor bilinears. Defining fermionic anticommutators
as \be \{ g, \tilde g \} \equiv - i g^T C \G^M \tilde g \pa_M
\label{anticomm} \, , \ee where $C = B \G^0$ is the charge
conjugation matrix, we find the following superalgebra: \bea
\{g_{m \over 2}^{\a a}, g_{m \over 2}^{\b b}\} &=&
\e^{\a\b}\e^{ab} l_m  , \nonu \{g_{m \over 2}^{\a a}, g_{-{m \over
2}}^{\b b} \} &=& \e^{\a\b}\e^{ab} l_0 + m \e^{ab} T^{\a\b} \, ,
\eea with \be T^{\a\b} = \left( \begin{array}{cc} -i j_+ & j_3 \\
j_3 & -i j_- \end{array} \right)\label{Ttensor} \, . \ee

\subsection*{M2-brane probes and their BPS properties}
Consider an M2-brane or anti-M2-brane probe, wrapped around the
S$^2$ and static with respect to the time $t$ in
(\ref{metricansatz}). Such a probe behaves as a point particle in
AdS$_3$ with action given in (\ref{sourceterms}). From (\ref{KV}),
we see the  $l_0$ Noether charge (eigenvalue $L_0$) is the energy
with respect to this time coordinate which  is given by \be L_0 =
{\p l \over l_3} \equiv Z.\label{noether} \ee

We shall show that such a probe  preserves half of the Killing
spinors (\ref{5DKspinor}). The supersymmetry condition is \be
\G_\k \e = \e \, , \ee where $\G_\k$ is the idempotent operator
that enters in the $\k$-symmetry transformations of the M2-brane
\cite{Bergshoeff:1987cm,Bergshoeff:1997kr}. In our case it is
given by \bea \G_\k &=& \pm i \G_{\hat t \hat \t \hat \f} \nonu
&=& \mp 2 i \g_{\hat z \hat{\bar z}} \otimes \g_{(6)} \, , \eea
where the upper (lower) sign corresponds to an (anti-)brane and
we have used the properties of our basis of gamma matrices in the
second line.
 Working out the SUSY condition a little further, we can write it as
\be \pm 2 \g_{\hat z \hat {\bar z}} \s_3 \varepsilon_0 =
\varepsilon_0.\label{probesusy} \ee This is a good  projection
condition and is compatible with the Majorana condition
(\ref{Maj5dKspin}); hence our probes leave 4 supersymmetries
unbroken, with the antibrane preserving the opposite  half of the
brane. In terms of our basis (\ref{Killspinorbasis}),  the
preserved Killing spinors are \bea g_{1 \over 2}^{\a +},\ g_{-{1
\over 2}}^{\a -} & \qquad &{\rm M2-brane,}\nonu g_{1 \over 2}^{\a
-},\ g_{-{1 \over 2}}^{\a +} & \qquad &{\rm anti-M2-brane.}
\label{probeKS} \eea The 1/2-BPS projection correlates the $L_0$
eigenvalue with the internal chirality, but imposes no restriction
on the R-symmetry quantum number $J_3$. The reason for the latter
is that our S$^2$-wrapped branes are invariant under R-symmetry.

We now investigate the supersymmetry properties of our probes from
the point of view of the worldvolume superalgebra. Our branes wrap
the nontrivial cycle  S$^2$ and carry a corresponding topological
charge. As usual, this charge enters as a central extension in the
superalgebra. We can compute the central terms using the results
of \cite{HackettJones:2003vz}: the anticommutators
(\ref{anticomm}) are modified to \be \{ g, \tilde g \} = - i g^T C
\G^M \tilde g \pa_M  \pm T_{M2} \int_{S^2} \o_{g, \tilde g}. \ee
The closed two-forms $\o_{g, \tilde g}$ can be computed by
evaluating the spinor bilinears \be \o_{g, \tilde g} = g^T C
\G_{\t \varphi} \tilde g\, d\t d\varphi. \ee Doing this we find
the extended superalgebra \bea \{g_{m \over 2}^{\a a}, g_{m \over
2}^{\b b}\} &=& \e^{\a\b}\e^{ab} l_m \, ,\nonu \{g_{m \over 2}^{\a
a}, g_{-{m \over 2}}^{\b b} \} &=& \e^{\a\b}\e^{ab} l_0 + m
\e^{ab} T^{\a\b} \mp m Z \e^{\a\b} \s_1^{ab} \, , \eea where the
(plus) minus sign corresponds to a (anti-)  brane.  This result
agrees with \cite{Gaiotto:2004pc}, where it was derived from a
construction of the worldvolume quantum mechanics. As discussed
there, the short multiplets saturate a BPS bound \be L_0 \geq j +
Z \ee and preserve $g_{1 \over 2}^{\a +},\ g_{-{1 \over 2}}^{\a
-}$ (for an M2-brane) or  $g_{1 \over 2}^{\a -},\ g_{-{1 \over
2}}^{\a +}$ (for an anti-M2-brane). This is consistent with the
analysis above, and in particular we see that the Noether charge
(\ref{noether}) is a consequence of the BPS bound for $j=0$.

\subsection*{The $\T$ nonconstant case: backreacted probes}
Consider the case where $\T$ is not constant, where we showed that
the local geometry becomes that of G\"{o}del space. To check the
supersymmetry of these solutions, we have to keep careful track
of the additional terms in the susy variation (\ref{susyvar})
proportional to derivatives of $\T$. For example,  the spin
connection receives extra contributions compared to
(\ref{spinconn}) \be \o = \o_{\eqref{spinconn}} + \a. \ee One
finds for the components of $\a$: \be
\begin{array}{lcl}
\a^{\hat t \hat z} = - { i \bar \pa \bar \T e^{-\f} \over 3 \T_2} ( dt + \chi) , &\qquad & \a^{\hat \varphi \hat z} = - { i \bar \pa \bar \T e^{-\f} \over 3 \T_2} \sin \t d \varphi ,\\
\a^{\hat z \hat{\bar z}} = { i \over 3 \T_2} (\pa \T dz + \bar \pa \bar \T d \bar z) , &\qquad & \a^{\hat a \hat z}  =  { i \bar \pa \bar \T e^{-\f} \over 3 l\T_2^{2/3}} e^{\hat a}_a d z^a ,\\
\a^{\hat \t \hat z} = - { i \bar \pa \bar \T e^{-\f} \over 3 \T_2}
d\t  , &\qquad & \a^{\hat a \hat{\bar z}}  =  -{ i \pa  \T e^{-\f}
\over 3 l\T_2^{2/3}} e^{\hat a}_a d z^a ,
\end{array}
\ee and mutatis mutandis for the complex conjugated components.
Further $\T$-derivatives enter in the expression for $\chi$ in
(\ref{chisol}) and in the 4-form field strength (\ref{F4}).
Keeping track of all these terms and after tedious algebra, one
finds that the backgrounds with nonconstant $\T$ preserve Killing
spinors  precisely of the form (\ref{5DKspinor}) (with $\T_2$ now
of course a varying function), subjected to additional projection
conditions: \bea
\g_{\hat { z}} \varepsilon^1_0 &=& 0  , \nonu\\
\g_{ \hat {\bar {z}}} \varepsilon^2_0 &=& 0.\label{eq:proj11D}
\eea These conditions project to one half of the supersymmetries,
leaving 4 local Killing spinors. This is in precise agreement with
our earlier probe analysis: we can combine the projections as \be
 2 \g_{\hat z \hat{\bar z}} \s_3 \varepsilon_0 = \varepsilon_0.
\ee In terms of our basis (\ref{KSmain}), the preserved Killing
spinors are $g_{1 \over 2}^{\a +},\ g_{-{1 \over 2}}^{\a -}.$
Comparing with (\ref{probeKS}) we see that our backgrounds
preserve the same supersymmetries as an S$^2$-wrapped M2-brane
probe and can hence be seen as backreacted geometries produced by
such probe configurations.

\section{Embedding in 5D $\caln$=1 supergravity\label{app-5Dsusy}}

We can interpret our system as a solution to $\caln=1$ supergravity in five
dimensions. We reduce the eleven-dimensional setup \eqref{Maction} to
five dimensions on the Calabi-Yau threefold, giving rise to
five-dimensional supergravity coupled to vector multiplets and one
hypermultiplet (see for instance \cite{Bergshoeff:2004kh,
Bellorin:2006yr}).  We will also briefly sketch how the
supersymmetry analysis is done in this setup, the details being
analogous to the calculation of the previous section in eleven
dimensions.

\subsection*{Reduction to five dimensions}

Starting from the eleven-dimensional action \eqref{Maction}, we
can perform a reduction over the Calabi-Yau threefold only, using
the ansatz \eqref{11Dreduction}. We are left with an action over
five spacetime dimensions of the form \bea
\frac{S_{5d}}{2\pi}&=&\frac{1}{l_{M}^3}\int dx^5\,\sqrt{-g}\left(R
-\frac{1}{2}\frac{\partial_\m \tau\partial^\m \bar \tau}{\tau_2^2}\right)\nonumber\\
&&-\frac{1}{2 l_M}\int G_{AB} F^A\wedge\star
F^B+\frac{D_{ABC}}{6}\int A^A\wedge F^B\wedge F^C\, . \eea In the
reduction, we have kept the five dimensional metric, but we only
allow the size $\tau_2$ of the internal manifold to vary,  while
the (internal) Calabi-Yau geometry is kept fixed.
This leads to constant vector multiplet scalars $J^A$ in five dimensions, hence they do not appear in the action. The two-form field strengths $F^A$ 
are  given in terms of the K\"ahler moduli $J^A$ as:
\begin{align}
  F^A &= \frac{l }{2 l_M } J^A \sin\theta d\theta\wedge d\phi\,,\label{eq:BG_Fstr}
\end{align}
Note that the complex scalar $\tau$ is part of the universal
hypermultiplet, for more information about the associated
geometry, see \cite{Strominger:1997eb}.

We have shown how our setup is described in the framework of
five-dimensional  $\caln=1$ supergravity, see for instance
\cite{Bergshoeff:2004kh,Bellorin:2006yr} for a general discussion of its
supersymmetric solutions. Moreover, as shown in the bulk of the
paper, we have obtained a non-trivial solution of this system,
with some of the hyperscalars turned on. Even though the general
form of these solutions has been discussed before, not many
explicit solutions with non-trivial hyperscalars are known.

\subsection*{Supersymmetry analysis}
Consider the supersymmetry variations in five dimensions, as can
be found in \cite{Bellorin:2006yr}. There are three equations
which will concern us, from the variation of the gravitino, the
gauginos and the hyperinos respectively. The gaugino variation is
identically zero, due to the constant scalars and the special
geometry properties. We are left with (all indices are flat):
\begin{align}
\label{eq:gravitino}
 0&= \delta_\epsilon \psi^i_M  = D_M \epsilon^i -
\frac{i}{8}\,Y_A F^{A\,PQ}\big(\Gamma_{MPQ} -
4\eta_{MP}\Gamma_Q\big)\,\epsilon^i
\\
\label{eq:hyperino} 0&=\delta_\epsilon \zeta^A  =
if_X^{iA}\Gamma^M\partial_M q^X\epsilon_i,
\end{align}
where $\psi^i_M$ are the gravitini, $\zeta^A$ the hyperini and
$\Gamma$ are representations of the Clifford algebra in five
dimensions. Note that the covariant derivative $D_M$ includes an
important term from the quaternionic $SU(2)$ connection. Also, our
gamma matrices are $i$ times the ones in \cite{Bellorin:2006yr} to
take into account the signature change.  We will use the equations
above to look for BPS solutions.

Demanding the hyperino variation to vanish, leaves us with a
projection condition making the solution 1/2-BPS. This projection
condition is identical to \eqref{eq:proj11D}. The BPS equation
that follows from the gravitino equation is solved by our
background. We will prove these two assertions below.

\subsubsection*{Gravitino variation and BPS equations}
Using algebra similar to that in \secref{app-susy}, we see that
the five-dimensional solution corresponding to the direct product
of a sphere with constant radius and three-dimensional G\"odel
space, as given in
(\ref{11Dreduction},\ref{eq:tauw},\ref{godelUHPmain}), indeed
satisfies the gravitino equation. We conclude that we have a
1/2-BPS solution to the equations of $\caln=1$ supergravity in five
dimensions.

\subsubsection*{Hyperino variation and projection condition}
The supersymmetry variation of the hyperinos is given by
\eqref{eq:hyperino}. In general, the quaternionic manifold on which the hyperscalars
live, has $SU (2) \times Sp(n)$ holonomy, with $n$ the number of
hypermultiplets. This holonomy structure is reflected in the index
structure of the vielbein $f_X^{iA}$. In our case, only the
universal hypermultiplet is excited. In the universal
hypermultiplet $q^X, X = 1\ldots4n$, only the components
corresponding to $\tau$ are nonzero. The quaternionic vielbein is
given by \cite{Strominger:1997eb}:
\begin{equation}
  f = \begin{pmatrix}
        0   &-v\\\bar v &0
        \end{pmatrix}\,,\qquad  v = \frac{d\tau}{\tau -\bar \tau}\,.
\end{equation}
Then the hyperino equation gives rise to the following two
conditions:
\begin{align}
  0 = \frac{\slashed \partial \tau}{\tau -\bar \tau} \epsilon^1\,,\qquad0=\frac{\slashed \partial \bar \tau}{\tau -\bar \tau} \epsilon^2\,.
\end{align}
If we now use that our solution for $\tau$ is holomorphic, we are
left with two (different) projection conditions for
$\epsilon^{1},\epsilon^{2}$. These are the same projection
conditions as in \eqref{eq:proj11D}, implying that the solution we
have is  1/2-BPS.

\section{Israel junction conditions}\label{IsJuncAp}
In this appendix we shortly review the derivation of the Israel
junction conditions in a simple and transparent, but slightly
unconventional way. We will give up some of the covariance of the
original derivation \cite{Israel:1966rt}, but in return the
simplicity of our approach allows us to set up a formalism that is
directly extended to more general field theories. Although in
this appendix we choose an explicit set of convenient local
coordinates it should be straightforward to rewrite our derivation
in a fully covariant fashion.

\paragraph{Notation:}
Before we get to the matching conditions themselves we start by
introducing some notation and conventions.

Imagine a function $f(y)$ that is continous for all $y$ and smooth
for all $y$ except at $y=y_0$. Then $f'(y)$ is generically
discontinuous at $y=y_0$ and $f''(y)$ has a delta-function
singularity at $y=y_0$ proportional to \be \Delta f'\equiv
\lim_{y\rightarrow y_0{}_-} f'(y)-\lim_{y\rightarrow y_0{}_+}
f'(y). \ee
In this section and paper we will often find it useful
to represent such functions as follows. A function $f$ with
properties as above can formally always be written as \be
f(y)=f_+(y)\Theta(y-y_0)+f_-(y)\Theta(y_0-y) \, , \ee where $f_+$
and $f_-$ are smooth functions and $\Theta$ is the Heaviside step function. Continuity at $y=y_0$ implies \be
\Delta f\equiv \lim_{y\rightarrow y_0{}_-} f(y)-\lim_{y\rightarrow
y_0{}_+} f(y)=f_+(y_0)-f_-(y_0)=0.\label{defdel} \ee Furthermore, one then has
\bea
f'(y)&=&f'_+(y) \Theta(y-y_0)+f'_-(y)\Theta(y_0-y) \, ,\\
f''(y)&=&f''_+(y) \Theta(y-y_0)+f''_-(y)\Theta(y_0-y)+\Delta f'\,
\delta(y-y_0) \, , \eea with generically $\Delta
f'=f'_+(y_0)-f'_-(y_0)\neq0$.

\paragraph{Israel junction conditions:}
The Israel junction conditions put restrictions on the
composition of a spacetime by `gluing' two metrics along a
hypersurface in a continuous way, i.e such that the induced metric
on the hypersurface coincides for both metrics. If each of the
metrics itself is a solution of the equations of motion then so is
the composite spacetime, at least away from the hypersurface. On
the hypersurface there will however generically be a non-vanishing
singular term, as derivatives along a coordinate orthogonal to the
hypersurface become singular there. As argued by Israel
\cite{Israel:1966rt} one can cancel these singular terms in the
e.o.m by introducing appropriate energy momentum also localized on
the gluing hypersurface. This added energy-momentum then has a
natural interpretation as the presence of a domainwall.

Let us derive the explicit form of the singular terms in the
equations of motion. Locally we can always choose to write the
metric under consideration in a set of Gaussian coordinates: \be
ds^2= N^2 dy^2+h_{ab}dx^adx^b \label{Gauss} \, . \ee These
coordinates are chosen such that the hypersurface along which the
two `bulk' metrics will be glued, is simply given by $y=y_0$. By
working in this specific set of coordinates we give up some of the
full diffeomorphism invariance, but it improves the clarity of our
discussion. We assume the components of the metric (\ref{Gauss})
to be made up of two different parts, coinciding at $y=y_0$: \be
N=N_+\Theta(y-y_0)+N_-\Theta(y_0-y) , \qquad
h_{ab}=h_{ab}^+\Theta(y-y_0)+h_{ab}^-\Theta(y_0-y) \, . \ee
Although we assume the metric to be `continuous', i.e. $\Delta
h_{ab}=\Delta N=0$, it doesn't necessarily have to be smooth in
$y$ at $y=y_0$.

Due to the second order nature of the Einstein equations it is
clear that the Einstein tensor of such a `glued' metric will be
singular. It is a straightforward exercise to check that the
Einstein tensor takes the form \be
G_{\m\n}=G_{\m\n}^+\Theta(y-y_0)+G_{\m\n}^-\Theta(y_0-y)+G_{\m\n}^{(s)}\delta(y-y_0)
\, , \ee with \bea
G_{ab}^{(s)}&=&-\frac{1}{N}\Delta\left(K_{ab}-h_{ab}K\right) ,\\
G_{ay}^{(s)}&=&0 , \\
G_{yy}^{(s)}&=&0 , \eea where $K_{ab}$ is the extrinsic curvature
and $K$ its trace. It is clear that the two glued metrics each
have to be solutions to the `bulk' equations of motion for the
composed metric to satisfy these equations; i.e.\,we also assume
the energy momentum tensor takes the form \be
T_{\m\n}=T_{\m\n}^+\Theta(y-y_0)+T_{\m\n}^-\Theta(y_0-y)+T_{\m\n}^{(s)}\delta(y-y_0)\,.
\ee Then the equations of motion at $y>y_0$ or $y<y_0$ take the
form \be G_{\m\n}^\pm=T_{\m\n}^\pm \, . \ee However, as we derived
above, the `gluing' introduces a singular part to the Einstein
tensor that still needs to be cancelled in the e.o.m. This is
accomplished by having an appropriate $T_{\m\n}^{(s)}$, which can
be interpreted as a domainwall with the correct effective tension.
More precisely one can add to the Lagrangian a term of the form:
\be L_{D.W.}=\delta(y-y_0)\tilde L_{D.W.} \, .\ee It is then clear
that the e.o.m coming from the standard Einstein-Hilbert
Lagrangian with the domainwall added demand that \be
\Delta\left(K_{ab}-h_{ab}K\right)=\frac{1}{\sqrt{-h}}\frac{\delta\tilde
L_{D.W.}}{\delta h^{ab}}\qquad\qquad\mbox{(at
$y=y_0$)}\label{IsJuncCond} \, . \ee This is our form of the
Israel-junction conditions.

\paragraph{Generalised junction conditions:}
Using the notation and formalism introduced above it is
straightforward to extend the Israel junction conditions to
more general field theories. Take any field theory with a
Lagrangian that only contains the fields and their first
derivatives and a domainwall source that only couples to the
fields and not their derivatives: \be
L_{\mathrm{total}}=L(\phi,\partial_{\m}\phi)+\delta(y-y_0)\tilde
L_{D.W.}(\phi) \, . \ee If we again choose Gaussian coordinates on
spacetime as in (\ref{Gauss}) and assume the fields to be
glued\footnote{More precisely:
$\phi=\phi_+\Theta(y-y_0)+\phi_-\Theta(y_0-y)$ with
$\Delta\phi=0$.} at $y=y_0$, then it is straightforward to derive
that the singular part is cancelled iff \be \Delta \frac{\delta
L}{\delta \partial_y \phi}=\frac{\delta\tilde L_{D.W.}}{\delta
\phi}\qquad\qquad\mbox{(at $y=y_0$)} . \label{GenJuncCond} \ee

Finally, note that the gravitational Israel-junction conditions
(\ref{IsJuncCond}) are a special case of this more general
formula. To see this one has to use an equivalent
 gravitational Lagrangian that only involves first derivatives of the metric:
\be L(g_{\m\n},\partial_\l
g_{\m\n})=\sqrt{-g}g^{\m\n}(\G_{\m\a}^\b\G_{\n\b}^\a-\G_{\m\n}^\a\G_{\a\b}^{\b}).
\ee One can then calculate that \be \frac{\delta L}{\delta
\partial_y h_{ab}}=-\sqrt{-h}(K^{ab}-h^{ab}K)\,. \ee Applying
formula (\ref{GenJuncCond}) then gives exactly the Israel-junction
conditions (\ref{IsJuncCond}).

\end{appendix}

\bibliographystyle{JHEP}
\bibliography{godel}
\end{document}